\documentclass[11pt]{article}
\usepackage[margin=1in]{geometry}
\usepackage{setspace}
\usepackage[round]{natbib}
\usepackage{multibib}

\newcites{App}{References for Appendix}

\usepackage{secdot}
\usepackage{epsfig}
\usepackage{amssymb}
\usepackage{amsmath}
\usepackage{amsthm}
\usepackage{color}
\usepackage{xcolor}
\usepackage{graphicx}
\usepackage{multirow}
\usepackage{enumerate}
\usepackage{subfig}
\usepackage{subcaption}
\usepackage{natbib}

 \bibpunct[, ]{(}{)}{,}{a}{}{,}%
\usepackage{algorithm,algorithmic}
\usepackage[normalem]{ulem}
\usepackage{float}
\usepackage{enumitem}
\usepackage{amsmath}
\usepackage{epstopdf}
\usepackage{graphicx}
\usepackage{tikz}
\usepackage{pgfplots}
\usepackage{subfig}
\pgfplotsset{compat=newest}
\usepackage{pgfplotstable}
\usetikzlibrary{arrows.meta,babel,calc,intersections,backgrounds,patterns,plotmarks,decorations.pathreplacing}
\usepackage{multirow}
\usepgfplotslibrary{fillbetween}
\tikzstyle{every picture} += [>=stealth]
\makeatother
\usepackage{lipsum}
\usepackage[justification=centering]{caption}
\usepackage{booktabs}
\usepackage[hyphens]{url}
\usepackage{mathtools}

\makeatletter
\newcommand\footnoteref[1]{\protected@xdef\@thefnmark{\ref{#1}}\@footnotemark}
\makeatother
\usepackage{apptools}
\usepackage{bbm}
\usepackage{chngcntr}
\usepackage{lipsum}
\usepackage{float}
\usepackage{hyperref}
\usepackage{bm}
\hypersetup{
    hidelinks,
    colorlinks=true,
    linkcolor=black,
    citecolor=blue
} 

\theoremstyle{plain}

\newtheorem{theorem}{\textbf{Theorem}}

\newtheorem{assumption}{\textbf{Assumption}}
\newtheorem{definition}{\textbf{Definition}}

\theoremstyle{definition}

\newcommand{\R}{\mathbb{R}}
\newcommand{\E}{\mathbb{E}}

\newcommand{\calP}{\mathcal{P}}

\newcommand{\softO}{\widetilde{O}}
\newcommand{\kmax}{k_{\max}}

\def\E{\mathbb{E}}

\def\P{\mathbb{P}}

\def\R{\mathbb{R}}

\newcommand{\X}{\mathcal{X}}

\usepackage{enotez}
\let\footnote=\endnote
\setenotez{backref = true}
\setenotez{list-name = {Endnotes}}
\usepackage[hang,flushmargin]{footmisc}
\makeatletter
\def\footnoterule{\relax
  \kern 1pt
  \hbox to \columnwidth{\vrule width 0.5\columnwidth height 0.5pt\hfill}
  \kern 1pt}
\makeatother

\tikzset{
 hatch distance/.store in=\hatchdistance,
 hatch distance=10pt,
 hatch thickness/.store in=\hatchthickness,
 hatch thickness=2pt
 }
 \makeatletter
 \pgfdeclarepatternformonly[\hatchdistance,\hatchthickness]{flexible hatch}
 {\pgfqpoint{0pt}{0pt}}
 {\pgfqpoint{\hatchdistance}{\hatchdistance}}
 {\pgfpoint{\hatchdistance-1pt}{\hatchdistance-1pt}}%
 {
 \pgfsetcolor{\tikz@pattern@color}
 \pgfsetlinewidth{\hatchthickness}
 \pgfpathmoveto{\pgfqpoint{0pt}{0pt}}
 \pgfpathlineto{\pgfqpoint{\hatchdistance}{\hatchdistance}}
 \pgfusepath{stroke}
 }
\makeatother
\usetikzlibrary{shapes.geometric, arrows, calc, shapes.misc, matrix, shapes, arrows, calc, positioning, shapes.geometric, shapes.symbols, shapes.misc, intersections, chains, scopes}
\tikzstyle{node1} = [circle, circle sides=6,minimum width=1.25cm, text badly centered, text width=2.35em,minimum height=1.15cm, draw=black, font=\small]
\tikzstyle{node2} = [rectangle,rounded corners, minimum width=1.25cm, minimum height=1cm, text centered, draw=black, font=\small]
\tikzstyle{arrow} = [thick,->,>=stealth]

\title{Beyond ATE: Multi-Criteria Design for A/B Testing}
\author{
  Jiachun Li\thanks{Authors are listed alphabetically.}\\
  Laboratory for Information and Decision Systems, MIT\\
  \texttt{jiach334@mit.edu}
  \and
  Kaining Shi\\
  Department of Statistics, University of Chicago\\
  \texttt{kainingshi@uchicago.edu}
  \and
  David Simchi-Levi\\
  Laboratory for Information and Decision Systems, MIT\\
  \texttt{dslevi@mit.edu}
}
\date{}

\begin{document}

\maketitle

\begin{abstract}
In the era of large-scale AI deployment and high-stakes clinical trials, adaptive experimentation faces a ``trilemma'' of conflicting objectives: minimizing cumulative regret (welfare loss during the experiment), maximizing the estimation accuracy of heterogeneous treatment effects (CATE), and ensuring differential privacy (DP) for participants. Existing literature typically optimizes these metrics in isolation or under restrictive parametric assumptions. In this work, we study the multi-objective design of adaptive experiments in a general non-parametric setting. First, we rigorously characterize the instance-dependent Pareto frontier between cumulative regret and estimation error, revealing the fundamental statistical limits of dual-objective optimization. We propose ConSE, a sequential segmentation and elimination algorithm that adaptively discretizes the covariate space to achieve the Pareto-optimal frontier. Second, we introduce DP-ConSE, a privacy-preserving extension that satisfies Joint Differential Privacy. We demonstrate that privacy comes ``for free'' in our framework, incurring only asymptotically negligible costs to regret and estimation accuracy. Finally, we establish a robust link between experimental design and long-term utility: we prove that any policy derived from our Pareto-optimal algorithms minimizes post-experiment simple regret, regardless of the specific exploration-exploitation trade-off chosen during the trial. Our results provide a theoretical foundation for designing ethical, private, and efficient adaptive experiments in sensitive domains.
\end{abstract}

\vspace{-2mm}

\section{Introduction and Literature Review}\label{sec:intro}

In the era of large-scale AI deployment, systems often function as ``black boxes,'' where complex, non-linear mappings make predicting real-world performance through mechanistic reasoning alone nearly impossible. This complexity is exacerbated by the unpredictability of human behavior. For instance, research on \textbf{algorithm aversion} \citep{dietvorst2015algorithm} reveals that users may lose total trust and reject superior AI recommendations after observing even a single minor, visible mistake. Conversely, prolonged exposure can lead to \textbf{deskilling}, as seen in clinical endoscopy where practitioners’ independent diagnostic abilities significantly decline after becoming reliant on AI assistance. These behavioral nuances necessitate evaluation frameworks that can quantify true \emph{causal impact} without making rigid assumptions about user responses.

The gold standard for such evaluation is the \textbf{Randomized Controlled Trial (RCT)}, or A/B testing. By randomly assigning users to variants, it identifies causal effects even when the underlying system is opaque. Clinical development offers a high-stakes analogue: like AI features, medical interventions involve significant uncertainty and safety risks. Thus, regulatory standards mandate ``adequate and well-controlled studies'' not merely to estimate an average effect, but to operationalize patient welfare and ensure robust decision-making under risk.

Despite the success and popularity of RCTs, there are some challenges and concerns from practice. \textbf{First, the ATE often masks substantial heterogeneity}.  By distilling an entire population’s response into a single number, the ATE often masks substantial heterogeneity that can determine the success or failure of an intervention.
In reality, a ``positive'' average can hide groups of people who are being actively harmed. Consider these diverse scenarios:
\begin{itemize}
    \item Digital Ecosystems: A new interface feature might significantly boost the productivity of ``power users'' who understand the system, while simultaneously confusing and driving away newcomers.
    \item Precision Medicine: A life-saving drug for one patient can be toxic to another based on their specific genotype, baseline risk, or pre-existing comorbidities.
\end{itemize}
This motivates the estimation of the \emph{conditional average treatment effect} (CATE), which is more difficult than estimating ATE, but also very important.

\textbf{Second, standard RCTs often incur significant welfare loss during the experiment}. Static randomization allocates a large fraction of participants to a potentially inferior variant until the trial concludes. In digital platforms, this manifests as degraded user experience, reduced engagement, and revenue loss; in high-stakes clinical domains, it results in avoidable risk to patients. The welfare loss incurred by not consistently selecting the optimal action is formalized as \emph{cumulative regret} in the bandit learning literature \citep{auer2002finite,lattimore2020bandit}. Negative experiences during experimentation can create lasting reputational harm or ethical backlash.
A stark illustration is the dynamic pricing experiment conducted by Amazon in 2000, where the company randomly varied the prices of DVDs to different users to estimate price elasticity. When users discovered they were paying different prices for the exact same item, the resulting backlash was severe, leading to accusations of price discrimination and a lasting erosion of consumer trust \citep{streitfeld2000amazon}.
This imperative is even more acute in trials for rare or fatal diseases, where the primary objective is to administer the most effective treatment possible to every participant.

To address these limitations, adaptive experimentation has emerged as a powerful framework to bridge the gap between static testing and real-time optimization. Historically, research and practice in this field have split into two distinct, well-established camps, each serving a different master:

\begin{itemize}
    \item Regret Minimization (The Bandit Paradigm): This approach, rooted in reinforcement learning, prioritizes participant welfare. By dynamically shifting resources toward the treatment that appears superior as evidence accumulates, bandit algorithms aim to minimize the ``regret'' or revenue loss incurred during the trial itself.
    \item Estimation Precision (The Neyman Paradigm): Conversely, this approach—often exemplified by Neyman Allocation—prioritizes statistical rigor. Its primary goal is to minimize the variance of the treatment effect estimate (ATE or CATE), ensuring that the final conclusions are as precise as possible for post-experiment deployment.
\end{itemize}
While both paradigms have been studied extensively, they are rarely considered in tandem. This creates a fundamental ``duality of purpose'' that leads to conflicting allocation strategies:
 Achieving the statistical precision required by a Neyman-style allocation typically necessitates broader exploration of all treatment options, including those that clearly appear suboptimal, while a ruthless focus on regret minimization restricts the algorithm's engagement with suboptimal arms, which in turn ``starves'' the exploration required for robust CATE estimation.
Despite their importance, these two objectives are usually optimized in isolation, leaving a critical gap in high-stakes environments—like clinical trials or large-scale AI deployment—where one cannot afford to sacrifice participant safety for data, nor data for safety.

\textbf{The third critical challenge in large-scale experimentation is the risk of privacy leakage.} In clinical trials, evaluating new interventions requires collecting sensitive data such as genetic markers and electronic health records. The disclosure of this information—whether through accidental leaks or sophisticated membership inference attacks—poses severe ethical and legal risks. For instance, in genomic studies, even aggregate statistics can be exploited to re-identify participants, potentially leading to genetic discrimination in insurance or employment \citep{homer2008resolving}. Similarly, in digital platforms, browsing histories and interaction logs can be deanonymized through linkage attacks with external datasets, revealing a user's private interests or political leanings \citep{narayanan2008robust}.

To mitigate these risks, Differential Privacy (DP) has become the gold standard, providing a formal guarantee that an individual's participation does not significantly alter the analysis outcome \citep{dwork2006calibrating}. While tech giants like Apple and Google have deployed DP for data collection \citep{erlingsson2014rappor}, it is well-known that \textbf{``privacy comes at a cost.''} In the context of online decision-making (e.g., contextual bandits), this cost is particularly high. Unlike offline estimation with independent data, online environments involve correlated samples, making the injection of noise far more complex. Our goal is to design an allocation mechanism that balances estimation accuracy and regret minimization under DP constraints. This requires a sophisticated calibration of noise to ensure privacy without catastrophically degrading the algorithm's performance.

The final, yet perhaps most consequential, concern is the \textbf{long-term impact of the experiment on the total population} following a full launch. Traditionally, the primary objective of an experiment is to identify the causal impact of treatments in order to design an optimal policy for deployment across a broader population. Consequently, it is equally, if not more, important to consider the \emph{post-experiment welfare}, such as the anticipated therapeutic outcomes in clinical practice or the expected revenue in a product ecosystem, when the selected policy is implemented at scale. A potential pitfall arises here: an over-aggressive focus on minimizing \emph{in-experiment} regret may lead to insufficient exploration of the action space, particularly for minority subgroups or rare clinical conditions. 
This creates a fundamental tension: does the effort to mitigate risk during the experiment through regret minimization lead to diminished post-experiment welfare due to ``starved'' exploration? Moreover, does the imposition of rigorous privacy constraints further exacerbate the welfare loss for the general population by masking the signals necessary for effective policy learning? 

The above questions necessitate a holistic evaluation framework that considers the entire life-cycle of an AI feature or medical intervention, from the initial randomized assignment to the eventual large-scale deployment. we provide a systematic theoretical foundation to analyze the optimization of different metrics in designing an adaptive experiment, and the trade-off between these objectives. We summarize our results as below.

\subsection{Main Contributions}
First, we characterize the instance-specific information-theoretic limits of CATE estimation subject to a constraint on maximum welfare loss. Departing from standard causal inference literature that focuses on i.i.d. estimation, and contextual bandit literature that prioritizes regret minimization, we tackle the complexity of \textbf{optimizing dual objectives}. We develop novel analytical techniques to rigorously establish the fundamental impossibility of simultaneously minimizing estimation error and regret.

To navigate this trilemma, we propose \textit{ConSE}, an adaptive experimentation framework based on sequential segmentation and elimination. This algorithm adaptively discretizes the covariate space to match the intrinsic complexity of the problem instance, achieving optimal estimation accuracy while strictly adhering to welfare constraints. Central to \textit{ConSE} is a tunable hyperparameter $\alpha$, which allows decision-makers to explicitly calibrate their tolerance for welfare loss. By establishing an algorithmic upper bound that matches the statistical lower bound, we demonstrate that \textbf{ConSE fully characterizes the achievable instance-dependent Pareto frontier}. Our analysis further reveals several fundamental insights into the nature of optimal design. First, we show that the two conventional extremes—static RCTs and pure regret minimization—are both strictly suboptimal in this dual-objective landscape. This underscores the necessity of a delicate balance between exploration and exploitation. Second, we find that any Pareto-optimal experiment naturally promotes group fairness in welfare loss, even without the imposition of explicit fairness constraints. Finally, addressing the reality that the true Pareto frontier is instance-dependent and initially unknown, \textit{ConSE} is designed to identify decisions that remain Pareto-optimal across all statistically indistinguishable instances, ensuring robust optimality in practice.

Second, we introduce \textit{DP-ConSE}, a privacy-preserving extension of our framework designed to protect both participant features and treatment responses from adversarial attacks. Implementing Differential Privacy (DP) in bandit settings is notoriously challenging due to the adaptive nature of data collection. Unlike prior works that rely on restrictive parametric assumptions, such as linear reward functions (\citealt{hanna2022differentially}, \citealt{shariff2018differentially}, \citealt{zheng2020locally}, \citealt{chen2022privacy}), we address the setting where such assumptions are risky, such as in clinical trials, by maintaining minimal assumptions on the CATE structure across patient types. This necessitates mechanisms distinct from existing literature.To the best of our knowledge, this work is the first to simultaneously address regret minimization and private causal effect estimation in a DP framework. Remarkably, we show that this privacy guarantee is achieved essentially ``for free,'' in the sense that \textbf{it imposes negligible additional cost on both estimation accuracy and regret} compared to the non-private baseline.

Finally, we systematically analyze the impact of experimental design on post-experiment welfare. Counter-intuitively, we establish that for \emph{any} choice of parameters within our Pareto-optimal, privacy-preserving algorithm \textit{DP-ConSE}, the resulting policy achieves the minimum possible welfare loss during full-scale deployment, regardless of the specific welfare-loss preference chosen during the experiment. The insight driving this result is that our rigorous characterization of the dual objectives not only provides a delicate trade-off among privacy, in-experiment loss, and estimation accuracy, but also ensures robustness for the final rollout. Consequently, decision-makers can optimize for experimental constraints without compromising the potential efficacy of the policy after full launch.

The remainder of the paper is organized as follows. First, we have a brief  review of related literature in Section \ref{sec:literature}. Section \ref{sec:formulation} introduces the formal definitions and formulation of the multi-objective optimization problem.  
Subsequently, in Section \ref{sec:lowerbound}, we characterize the statistical limit of estimating CATE under a regret constraint for each specific instance. We then present ConSE in Section \ref{sec:ConSE}, an instance-adaptive optimal algorithm that appropriately calibrates the exploration rate to balance this trade-off.
Then in Section \ref{sec:privacy_simple_regret}, we provide \textit{DP-ConSE} that protects the sensitive information of the individuals and show that it does not iccur significant error in causal effect estimation or more welfare loss.
We also prove that for any choice of hyper-parameter $\alpha$ in \textit{DP-ConSE}, we can output a treament policy that is instance-dependent optimal. We conclude in Section \ref{sec: conclusion}.  

\subsection{Literature Review}\label{sec:literature}
\noindent\textbf{Adaptive Experimental Design.}
Experimental design has witnessed a surge in popularity across operations research, econometrics, and statistics (see, e.g., \citealt{johari2015always}, \citealt{bojinov2021panel}, \citealt{bojinov2023design}, \citealt{xiong2023optimal}). Adaptive experimental design emerges as a particularly relevant area to our current focus (\citealt{hahn2011adaptive}, \citealt{atan2019sequential}, \citealt{greenhill2020bayesian}). Multi-armed bandits (MAB) can also be viewed as a form of adaptive experimental design, albeit with the primary objective of minimizing regret, whereas most literature on adaptive experimental design primarily concentrates on (conditional) average treatment effects (ATE).  
\cite{kato2020efficient} investigate adaptive experiments for ATE under observable covariates.  \cite{dai2023clip} introduce a metric called Neyman regret and show that an adaptive design achieving asymptotically optimal variance corresponds to sublinear Neyman regret, thus framing the problem as one of regret minimization. Similarly, \cite{zhao2023adaptive} consider a comparable setting but employ a competitive analysis framework.  \cite{li2024optimal} extends this framework to the case with covariate and prove a non-asymptotic semi-parametric efficiency bound for general experimental design. 

Another closely related line of adaptive experimentation with heterogeneous covariate is the regret minimization of contextual bandit.
One widely considered model is contextual linear bandit, where the reward function is assumed to have a linear structure \citep{abbasi2011improved}. Extensive algorithms have been developed, mostly based on variants of Lin-UCB \citep{chu2011contextual, li2017provably} and achieves $O(\sqrt{T})$ minimax regret  or $O(\log T/\delta )$ instance dependent regret \citep{lattimore2017end, hao2020adaptive}. Such parametric assumption is typically considered risky and restrictive in a causal effect estimation scenario, and nonparametric framework is more widely considered. In this case, the minimax optimal regret will no longer be 
$O(\sqrt{T})$ but depends on the complexity of the function class \citep{foster2020beyond, simchi2022bypassing}, mostly the smoothness of the outcome function.
In this work, we only assume the Lipschitz continuity of  outcome function, which is arguably the most widely adopted, minimal assumption \citep{rigollet2010nonparametric}. To characterize the hardness of each instance, the margin condition is the most well-known condition that is widely adopted in most nonparametric contextual bandit literature \citep{perchet2013multi,hu2020smooth}.

\noindent \textbf{Multi-objective (Contextual) Bandit Optimization.}
Another emerging area is multitasking bandit problems, where minimizing regret is not the sole objective (see, e.g., \citealt{yang2017framework}, \citealt{yao2021power}, \citealt{zhong2021achieving}).  
\cite{erraqabi2017trading} also investigate the trade-off between regret and estimation error, proposing a novel loss function that jointly captures these two objectives. The work most closely related to this paper is \cite{simchi2023multi}, which examines the trade-off between regret and ATE estimation.
Following this work, a sequence of works try to extend to more general settings, including multiple treatments \citep{zuo2025pareto},  best arm identification \citep{zhong2021achieving}, network interference \citep{zhang2024online} and contextual linear bandit \citep{duan2024regret}.   
We extend their framework to the nonparametric contextual setting, which is more natural in the A/B testing framework. Moreover, most existing work aims at characterize \textbf{worst case optimality}, while we in this work shows the minimal possible estimation error and regret  \textbf{for every instance}, which is much stronger since while the Pareto optimality in previous works can not be dominated under the worst case, it can be \textbf{over-pessimistic in most regular cases}, and we show that indeed in most cases we can dominate both better accuracy as well as regret compared to the worst case Pareto optimality in previous works. The technique required to improve from worst case optimality to instance-dependent optimality requires novel techniques, careful analysis for the complexity of each instance, as well as a general enough measure that covers both the margin condition from the regret minimization part, and the statistical complexity from the causal estimation part.

\noindent \textbf{Differentially Private (Contextual) Bandit Learning and Causal Estimation.}
Differential privacy (\citealt{dwork2006calibrating}) has emerged as the gold standard for privacy-preserving data analysis, ensuring that the output of an algorithm depends minimally on any single individual datum. Differentially private variants of online learning algorithms have been successfully developed in various settings (\citealt{guha2013nearly}), including private UCB algorithms for the multi-armed bandit (MAB) problem (\citealt{azize2022privacy}, \citealt{tossou2016algorithms}) as well as UCB adaptations in linear bandit settings (\citealt{hanna2022differentially}, \citealt{shariff2018differentially}).  
However, when it comes to differentially private (DP) contextual bandits, most works focus on contextual linear bandits (\citealt{shariff2018differentially}, \citealt{hanna2022differentially}, \citealt{charisopoulos2023robust}) and adopt relaxed notions such as \textit{joint-DP} or \textit{anticipating-DP}. These methods are generally variants of Lin-UCB (\citealt{abbasi2011improved}), which is suboptimal. It was until very recently that the optimal regret in contextual linear bandit problem with privacy constraint  is proved using novel confidence interval based $L_1$ analysis \citep{li2024optimal}. When it comes to nonparametric contextual bandit, the understanding of optimal regret and efficient algorithm are still very limited.

There has been some initial work on differentially private causal inference methods. \cite{lee2013neighborhood} proposed a privacy-preserving inverse propensity score estimator for estimating the average treatment effect (ATE). In \cite{kusner2016private}, the authors focus on privatizing statistical dependence measures, such as Spearman’s \(\rho\) and Kendall’s \(\tau\), aiming to derive privatized scores that still accurately infer the causal direction between two random variables. Furthermore, \cite{agarwal2021causal} study parametric estimation of causal parameters within the local differential privacy framework, and all the above works focus on statistical estimation with i.i.d. samples without consideration of adaptive experiments.

\section{Problem Formulation}\label{sec:formulation}
In adaptive experiment design with heterogeneous treatment effect, there is a binary set $\mathcal{A}=\{0,1\}$ of arms (i.e., treatments or controls) and a $d$-dimensional feature set $\mathcal{X}\subset \mathbb{R}^d$.  Suppose $n$ is the time horizon (or the total number of experimental units). At each time $t \leqslant n$, for every arm $a \in \mathcal{A}$ and feature of the unit $x \in \mathcal{X}$, we can observe a reward (outcome) $Y_t$. The random covariate $X_t$ is drawn independently from a fixed distribution $P_X$ on the hypercube $\mathcal{X} = [0,1]^d$.
After observing feature $X_t\in \mathcal{X}$,
a treatment allocation policy $\pi$ selects an arm $a_t(X_t) \in \{1, 2\}$ based on past observations $H_{t-1}:= \{X_s, a_s, Y_s\}_{s=1}^{t-1}$ and the current covariate $X_t$.
Then a reward $Y_t = Y_t^{(a_t)}$ is observed. The reward from arm $i \in \{1,2\}$ is a random $ Y_t^{(i)}$ variable bounded by $[0,1]$ with the expectation of $f^{(i)}(X_t)$.
The pair of mean reward functions $(f^{(1)}, f^{(2)})$ is unknown, and we only assume that it's Lipschitz continuous.  Formally, it's defined as:
\begin{align}
    |f^{(i)}(x) - f^{(i)}(x')| \leqslant L\|x - x'\|, \quad \forall x, x' \in \mathcal{X}, i=1,2
\end{align}
for some $L > 0$.
Such assumption is also broadly adopted and considered minimal in causal inference literature in order to have proper estimation of conditional outcome and treatment effect \citep{wager2024causal}. 
We define the conditional average treatment effect (CATE) of a feature $x$ as $\Delta_f(x):=f^{(2)}(x)-f^{(1)}(x)$, for any $X \in \mathcal{X}$. 
Denote all possible distributions satisfying the mentioned assumptions to constitute a feasible set $\mathcal{E}_0$.
As described in Section \ref{sec:intro}, we are interested in designing a policy $\pi$ with the following objectives:
\begin{enumerate}
    \item \textbf{Minimize Cumulative Regret}: The expected cumulative regret is
  \begin{equation}\label{eq:cum_regret}
      R_n(\pi) = \E \left[ \sum_{t=1}^n \left( f^{(\pi^*(X_t))}(X_t) - f^{(a_t)}(X_t) \right) \right],
  \end{equation}  
    where $\pi^*(x) = \arg\max_i f^{(i)}(x)$ is the oracle optimal policy.
    \item \textbf{Minimize Estimation Error}: After $n$ rounds, an estimator of CATE $\hat{\Delta}(X)$ maps the history $\mathcal{H}_n:= \{X_t, a_t, Y_t\}_{t=1}^n$ to an estimation of $\Delta(X)$
    . The estimation error is
    \begin{equation}\label{eq:error}
        E_n(\hat{\Delta}) = \E_X \left[ \|\hat{\Delta}(X) -\Delta(X)\|_2^2 \right].
    \end{equation}
    \item \textbf{Minimize Post-Experiment Regret}:
After $n$ rounds, the experimenter outputs a fixed policy $\pi': \mathcal{X} \to \{1,2\}$ based historical observation $\mathcal{H}_n:= \{X_t, a_t, Y_t\}_{t=1}^n$. The post-experiment regret of $\pi'$ on the population is 
\begin{equation}\label{eq:simple_Regret}
    r(\pi')=\E_X\left[f^{(\pi^*(X))}(X)-f^{(\pi'(X))}(X)\right].
\end{equation}
\end{enumerate}
A design of adaptive experiment can then be represented by an admissible policy-estimator pair $(\pi, \hat{\Delta})$.
Different from classical design of A/B testing, which aims at minimize estimation error of CATE, or design of contextual bandit algorithm which tries to minimize  cumulative loss, the optimal design of adaptive experiment in this paper is solving the following multi-objective optimization problem for every instance $\nu$:
\begin{equation}
\min _{(\pi, \hat{\Delta})} \left(\mathcal{R}_{n,\nu}( \pi), E_{n,\nu}\left( \hat{\Delta}\right)\right)
\end{equation}
where we use the subscript $\nu$ to denote the covariate and outcome distribution. Eq. (1) mathematically describes the two goals: minimizing both the regret the estimation error. 

In this work, we focus on the asymptotic scaling of regret and estimation error with respect to the time horizon $n$. To facilitate a rigorous comparison, we introduce the following order-wise notation. For two positive functions $f(n)$ and $g(n)$, we denote:
\begin{itemize}
    \item $f(n) \succeq g(n)$ if $g(n) \leqslant \widetilde{\mathcal{O}}(f(n))$;
    \item $f(n) \prec g(n)$ if $f(n) = o(g(n)n^{-\alpha})$ for some constant $\alpha > 0$;
    \item $f(n) \asymp g(n)$ if $f(n) = \widetilde{\Theta}(g(n))$.
\end{itemize}

We characterize the performance of an adaptive experimental design, represented by a policy-estimator pair $(\pi, \hat{\Delta})$, through its feasible performance set across the instance class $\mathcal{E}_0$. Formally, we define the \textit{performance frontier} of $(\pi, \hat{\Delta})$ as:
\begin{align}
    \mathcal{F}(\pi, \hat{\Delta}) = \left\{ \left(\mathcal{R}_{n,\nu}(\pi), E_{n,\nu}(\hat{\Delta})\right) \mid \nu \in \mathcal{E}_0 \right\}.
\end{align}

\begin{definition}[Pareto Dominance]
    Consider an instance $\nu$ and two policies $(\pi_1, \hat{\Delta}_1)$ and $(\pi_2, \hat{\Delta}_2)$. We say that $(\pi_1, \hat{\Delta}_1)$ \textbf{Pareto dominates} $(\pi_2, \hat{\Delta}_2)$ with respect to $\nu$ if there exist constants $M > 1$ and $C_1, C_2, \gamma_1, \gamma_2 > 0$ such that, for all $n \geqslant M$, the following holds:   
    \begin{align*}
        \begin{cases} 
            \mathcal{R}_{n,\nu}(\pi_1)\leqslant C_1 (\log n)^{\gamma_1}\cdot\mathcal{R}_{n,\nu}(\pi_2) \\ 
            E_{n,\nu}(\hat{\Delta}_1)<\frac{E_{n,\nu}(\hat{\Delta}_2)}{C_2(\log n)^{\gamma_2}}
        \end{cases}
        \text{ or }
        \begin{cases} 
            \mathcal{R}_{n,\nu}(\pi_1)<\frac{\mathcal{R}_{n,\nu}(\pi_2)}{C_2(\log n)^{\gamma_2}} \\ 
            E_{n,\nu}(\hat{\Delta}_1)\leqslant  C_1 (\log n)^{\gamma_1}\cdot E_{n,\nu}(\hat{\Delta}_2)
        \end{cases}
    \end{align*}
\end{definition}

A policy-estimator pair $(\pi_1, \hat{\Delta}_1)$ is said to be \textbf{Pareto-superior} to $(\pi_2, \hat{\Delta}_2)$ if for every performance outcome of the latter, the former can achieve a strictly better or equal outcome order-wise. Specifically, we require that for all $(R_2, e_2) \in \mathcal{F}(\pi_2, \hat{\Delta}_2)$, there exists $(R_1, e_1) \in \mathcal{F}(\pi_1, \hat{\Delta}_1)$ such that $R_1 \preceq R_2$ and $e_1 \preceq e_2$.

Having established the framework for the regret-estimation trade-off, we now address the second core inquiry of this work: quantifying the cost of privacy protection in adaptive experimentation. 
To rigorously analyze this trade-off, we must first adopt an appropriate definition of privacy for sequential decision-making. The standard notion of Differential Privacy (DP) is often too restrictive in bandit settings. Instead, we adopt the concept of \textit{Joint Differential Privacy} (JDP), originally proposed by \cite{shariff2018differentially} for linear contextual bandits and subsequently adapted as \textit{Anticipating Differential Privacy} (ADP) by \cite{chen2022privacy}.
\begin{definition}[Joint Differential Privacy / Anticipating DP] \label{def:JDP-interactive}
An adaptive algorithm $\pi$ preserves \emph{$\varepsilon$-Joint Differential Privacy} (or equivalently, \emph{Anticipating DP}) if, for any time horizon $n$, and any two neighboring datasets $\mathcal{D}=\{(X_s,Y_s)\}_{s=1}^n$ and $\mathcal{D}'=\{(X_s',Y_s')\}_{s=1}^n$ that differ only at a single round $t$ (i.e., $(X_t, Y_t) \neq (X_t', Y_t')$), the following inequality holds for the sequence of actions excluding the user at time $t$:
\begin{align*}
    \mathcal{P}^{\pi}\left( a_{-t} \in E \mid \mathcal{D} \right) \leqslant e^{\varepsilon} \mathcal{P}^{\pi}\left( a_{-t} \in E \mid \mathcal{D}' \right) + \delta, \qquad \forall E \subseteq \mathcal{A}^{n-1},
\end{align*}
where $a_{-t} = \{a_1, \dots, a_{t-1}, a_{t+1}, \dots, a_n\}$ denotes the sequence of actions for all users except user $t$. The probability measure $\mathcal{P}^{\pi}$ accounts solely for the internal randomness of the policy $\pi$, defined explicitly as:
\begin{align*}
   \mathcal{P}^{\pi} \left(a_1,\dots,a_n \mid \mathcal{D} \right) = \prod_{s=1}^n \pi_{s}(a_{s} \mid \{(X_j, a_j, Y_j)\}_{j < s}, X_{s}).
\end{align*}
\end{definition}
This definition diverges from classical Differential Privacy by excluding the action $a_t$ observed by the $t$-th user from the privacy guarantee. This relaxation—termed Anticipating DP (ADP) by \cite{chen2022privacy}—is critical for two primary reasons:

\begin{enumerate}
    \item \textbf{Avoidance of Linear Regret:} Enforcing classical DP, which would require the sequence of \emph{all} actions (including $a_t$) to be insensitive to the user's own private input $X_t$, fundamentally precludes the possibility of learning. If $a_t$ cannot depend on $X_t$, the algorithm cannot personalize treatments, inevitably leading to linear regret.
    \item \textbf{Causal Consistency:} In an adaptive experiment, the algorithm's decisions prior to time $t$ are causally independent of the $t$-th user's arrival. Consequently, the private information $(X_t, Y_t)$ cannot influence the history $a_{1}, \dots, a_{t-1}$. An adversary attempting to infer the sensitive attributes of user $t$ can strictly leverage only the \emph{downstream} allocations $a_{t+1}, \dots, a_n$, which may depend on the updated model state. ADP captures precisely this threat model: it ensures that the system's future behavior does not leak information about a specific user's past contribution.
\end{enumerate}
For a comprehensive analysis of the theoretical properties of ADP, we refer the reader to \cite{chen2022privacy}.

Following the privacy definition for the adaptive allocation process, we must also rigorously define the privacy requirement for the final output of the experiment: the CATE estimator itself. While the experimental trajectory is protected under JDP, the final estimator $\hat{\Delta}$—which is typically released to the public or policy-makers—must satisfy the standard notion of differential privacy to prevent information leakage from the trained model parameters.

\begin{definition}[Differentially Private CATE Estimator] \label{def:DP-estimator}
Let $\mathcal{M}$ be a randomized estimation algorithm that maps a dataset $\mathcal{D}=\{(X_t, A_t, Y_t)\}_{t=1}^n$ to a CATE estimator function $\hat{\Delta}: \mathcal{X} \to \mathbb{R}$. The algorithm $\mathcal{M}$ is said to be \emph{$(\varepsilon, \delta)$-differentially private} if for any two neighboring datasets $\mathcal{D}$ and $\mathcal{D}'$ differing by a single individual, and for any measurable subset of possible estimators $\mathcal{S}$ in the range of $\mathcal{M}$, it holds that:
\begin{align*}
    \mathbb{P}\left( \mathcal{M}(\mathcal{D}) \in \mathcal{S} \right) \leqslant e^{\varepsilon} \mathbb{P}\left( \mathcal{M}(\mathcal{D}') \in \mathcal{S} \right) + \delta.
\end{align*}
\end{definition}

\noindent \textbf{Remark.} 
It is crucial to distinguish between Definition \ref{def:JDP-interactive} (JDP) and Definition \ref{def:DP-estimator} (Standard DP). The former governs the \emph{online interaction} during the experiment, ensuring that a participant's data does not influence the treatments assigned to other concurrent participants in a revealing way. The latter governs the \emph{offline release} of the learned policy or treatment effects. In our framework, \textit{DP-ConSE} satisfies JDP for the allocation sequence $a_{1:n}$ and Standard DP for the final estimator $\hat{\Delta}$, thereby providing end-to-end privacy protection.

\section{Statistical Limit of CATE Estimation with Regret Constraint}\label{sec:lowerbound}
In this section, we characterize the best  possible estimation accuracy for CATE in \eqref{eq:error} with the constraint of 
a maximum cumulative welfare or revenue loss during the experiment. We first have the following assumption for the regularity of the adaptive experimentation.

\begin{figure}
    \centering
    \includegraphics[width=0.8\linewidth]{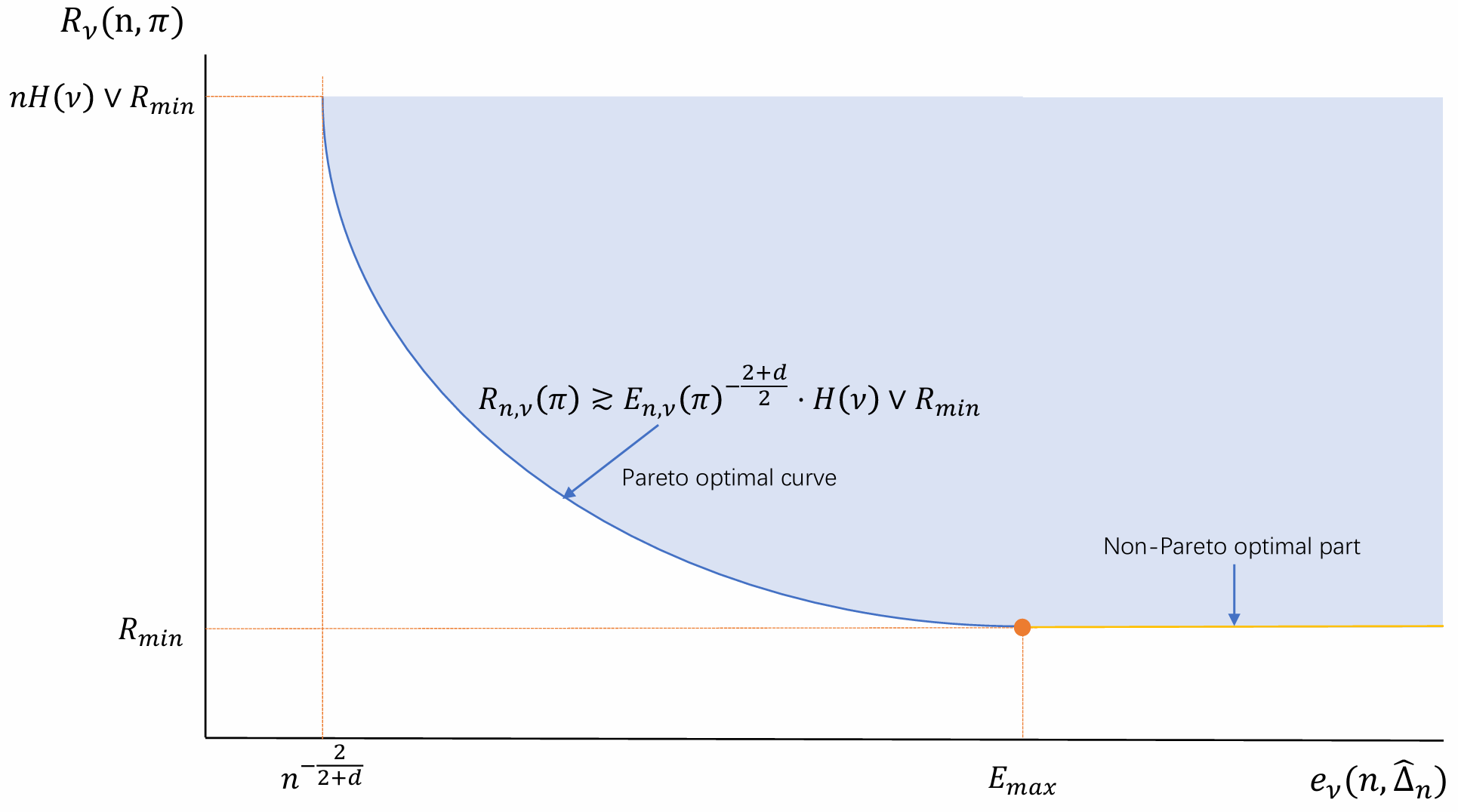}
    \caption{Fundamental statistical limit of the tradeoff between regret and CATE estimation}
    \label{fig:lowerbound}
\end{figure}

\begin{assumption}[No Regret Super-efficiency] \label{ass: minimal regret assumption}
    For any instance $\nu$, we assume the regret of the policy $\pi$ in $\nu$ is at least:
\begin{align*}
    R_{n,\nu}(\pi)\geqslant R_{\min}:= \softO\left( \E_X\left[\min\{n|\Delta(X)|, |\Delta(X)|^{-1-d}\}\right]\right)
\end{align*}
\end{assumption}
The aforementioned condition posits that the behavior of the adaptive experiment is \emph{regular}, implying that the algorithm conducts the minimum exploration necessary to distinguish the optimal arm prior to full exploitation. This regularity condition is satisfied by state-of-the-art regret minimization algorithms \citep{foster2020beyond, simchi2022bypassing} as well as modern adaptive A/B testing methods \citep{dai2023clip,zhao2023adaptive}. Furthermore, it aligns with established instance-dependent regret lower bounds in non-parametric contextual bandits under the margin condition. Specifically, consider an instance $\nu$ satisfying the margin condition with parameter $\alpha \geqslant 0$:
\begin{equation}
    \exists\ \delta_0 \in (0,1), D > 0, \quad \text{s.t.} \quad \forall \delta \in [0, \delta_0],\ \mathbb{P}_X\left(0 < |f^{(2)}(X) - f^{(1)}(X)| \leqslant \delta\right) \leqslant D\delta^{\alpha}.
\end{equation}
Existing literature establishes that the regret lower bound for instances satisfying this condition scales as \scalebox{0.8}{$\tilde{O}\left(n^{\frac{1-\alpha+d}{2+d}}\right)$} \citep{hu2020smooth}. We demonstrate that the lower bound required by our assumption,
\begin{align}\label{eq: minimal regret v.s. margin condition}
        \mathbb{E}_X\left[\min\{n|\Delta(X)|, |\Delta(X)|^{-1-d}\}\right] \leqslant \tilde{O}\left(n^{\frac{1-\alpha+d}{2+d}}\right),
\end{align}
is no more stringent than the standard margin-based lower bound for $\nu$ with any parameter $\alpha$. The proof shows in the appendix \ref{proof: minimal regret v.s. margin condition}. 
We now establish an exact characterization of the optimal causal effect estimation error subject to a cumulative regret constraint.
\begin{theorem} \label{thm-lower}Consider the problem setting defined above with $d \geqslant 1$ and $L > 0$. Define the local neighborhood $B(\nu) := \{\nu' : C_1 \leqslant \frac{\Delta_{\nu'}(X)}{\Delta_{\nu}(X)} \leqslant C_2 \text{ for all } X\}$ representing the set of instances with CATE profiles comparable to $\nu$ up to constants $C_1, C_2$. For any adaptive experimentation policy $\pi$ and instance $\nu \in \mathcal{E}$, there exists a proximate instance $\nu' \in B(\nu)$ such that the causal estimation error is lower-bounded by the cumulative regret:\begin{equation}\begin{aligned}R_{n,\nu'}(\pi) \geqslant \softO\left( \left(E_{n,\nu'}^{-\frac{2+d}{2}}(\pi) \cdot \E_X\left[|\Delta_{\nu'}(X)|^{\frac{2}{4+d}}\right]^\frac{4+d}{2}\right) \vee \E_X\left[\min\{n|\Delta(X)|, |\Delta(X)|^{-1-d}\}\right]\right).\end{aligned}\label{eq:lower_bound}\end{equation}\end{theorem}
Let $H(\nu) := \mathbb{E}_X\left[|\Delta_{\nu}(X)|^{\frac{2}{4+d}}\right]^{\frac{4+d}{2}}$ denote the coefficient characterizing the complexity of instance $\nu$. Theorem \ref{thm-lower} establishes the following bound for any adaptive policy:

\begin{equation}
R_{n,\nu'}(\pi) \geqslant \tilde{O}\left(\left(H(\nu)\cdot E_{n,\nu'}^{-\frac{2+d}{2}}(\pi)\right) \vee R_{\min}\right)
\end{equation}
This result succinctly quantifies how the optimal estimation accuracy $E_{n,\nu}(\pi)$ is constrained by the cumulative regret $R_{n,\nu}(\pi)$ and the instance complexity $H(\nu)$. Here, $H(\nu)$ acts as the ``exchange rate'' between welfare loss and statistical knowledge: 

\begin{itemize}
    \item \textbf{The High-Cost Regime:} When $H(\nu) \geqslant O(1)$, often due to significant treatment differences across a large population, precise estimation requires allocating many participants to the suboptimal group, severely impacting welfare.
    \item \textbf{Beyond the Worst Case:} While prior literature often focuses on global minimax limits, our analysis shows that for instances with a small $H(\nu)$, the regret required for high accuracy is substantially reduced.
\end{itemize}
An illustration of the statistical lower bound is provided in Figure \ref{fig:lowerbound}. Notably, once the cumulative regret reaches the threshold $R_{\min}$, it is no longer Pareto optimal to further sacrifice estimation accuracy, as doing so yields no further reduction in regret; consequently, an optimal experimental design should not operate in this redundant regime. However, it is important to emphasize that $R_{\min}$ is inherently \textbf{unidentifiable} within a finite time horizon. This leads to a scenario where multiple candidate instances—each characterized by distinct Pareto frontiers, minimal regret levels, and corresponding maximal estimation errors—remain indistinguishable to the learner. As a result, the entire theoretical Pareto frontier may not be uniformly achievable across all potential instances. In the subsequent section, we define and characterize the \textbf{achievable Pareto frontier}, which represents the subset of the frontier that maintains Pareto optimality across all indistinguishable instances simultaneously.

The proof of Theorem \ref{thm-lower} necessitates novel techniques distinct from standard non-parametric lower bound methods, such as Fano's or Assouad's method, which typically rely on constructing a static set of hard instances. In our setting, the adaptive nature and the regret constraint introduce significant new challenges. First, the construction of hard instances depends on the regret $R_n(\pi, \nu)$ of $\pi$ on $\nu$, which effectively serves as the ``effective sample size'' for the adaptive experiment.  Second, and perhaps most subtly, while we can identify a hard instance $\nu'$ yielding a large error relative to the regret on $\nu$ (i.e., $R_{n,\nu'}(\pi) \geqslant \softO(H(\nu) \cdot E_{n,\nu}^{-\frac{2}{2+d}})$), there is a \textbf{mismatch in the quantities being compared}. It is possible that the regret on $\nu'$ is significantly smaller than on $\nu$, effectively allowing the algorithm to ``cheat'' the trade-off locally (e.g., if the estimation error on $\nu$ is $0$ because the algorithm simply outputs a fixed function). Consequently, a direct comparison fails to prove that reducing estimation error \emph{fundamentally} mandates exploration and thus larger regret. 
 
To overcome this obstacle, we introduce a novel recursive argument. Starting from $\nu$, we identify a candidate $\nu^{(1)} \in B(\nu)$ with the desired error profile. If the regret of $\pi$ on $\nu^{(1)}$ is comparable to that on $\nu$, the proof concludes. If not, $R_n(\pi,\nu^{(1)})$ must be significantly smaller than $R_n(\pi,\nu)$. We then treat $\nu^{(1)}$ as a base instance to find a subsequent $\nu^{(2)} \in B(\nu^{(1)})$ with even larger estimation error (since the reduced regret implies a smaller effective sample size). We iterate this process to find $\nu^{(3)}, \nu^{(4)}$, and so on. In each step of the recursion, the estimation error increases while the regret decreases. Crucially, since the algorithm is assumed to be regular, it must incur a non-trivial minimal regret. Thus, the recursion is guaranteed to terminate in finite steps, yielding the desired instance. It is also worth noting that our construction partitions the feature space $\mathcal{X}$ into segments $\mathcal{X}^{(i)}:= \{X: 2^{-i}< \Delta(X) \leqslant 2^{-i+1}\}$. We demonstrate that a single ``hardest'' segment type determines the complexity of the instance. This finding directly motivates the adaptive segmentation algorithm, \textit{ConSE}, proposed in the subsequent section.

\section{ConSE: An Adaptive Experimentation Framework that Estimates Causal Effect with Minimal Welfare Loss}\label{sec:ConSE}

In this section, we introduce \textit{ConSE}, a sequential segmentation and elimination framework designed to estimate CATE while strictly controlling potential welfare and revenue loss.The high-level intuition of this framework is to leverage the heterogeneity of the treatment effect across the covariate space $\mathcal{X}$. Specifically, we aim to segment the population based on the \emph{magnitude} of the treatment effect (i.e., the ``signal strength'' of the new treatment relative to the baseline).Relying on the Lipschitz continuity of the outcome function, we partition the feature space $\mathcal{X}=[0,1]^d$ into a union of disjoint hypercubes (bins). The construction ensures that within each bin, the treatment effect scale is homogeneous: for any $X, X'$ belonging to the same bin, the ratio $\frac{\Delta(X)}{\Delta(X')}$ is bounded within $[m, M]$ for some constants. Crucially, the diameter of each bin is adapted to match the scale of the local treatment effect $\Delta(X)$. Mechanistically, \textit{ConSE} constructs this segmentation via a recursive dyadic splitting process. We initialize the entire space $\mathcal{X}$ as the root bin $B_0$. The process proceeds as follows: if the treatment effect magnitude in a bin satisfies $\Delta(X) \geqslant C$ for some constant threshold, the bin is retained without further splitting. Otherwise, we recursively bisect each dimension of the bin, generating $2^d$ sub-bins. For a bin at level $i$ with diameter $O(2^{-i})$, we continue this splitting until the local treatment effect scales with the diameter, i.e., $\Delta(X) \approx O(2^{-i})$. This adaptive discretization strategy ensures that we effectively manage only $O(\log n)$ types of discrete feature regions, indexed by the bin they occupy. The primary advantage of this multi-scale binning is that it establishes an optimal structure for information sharing: \begin{itemize}\item \textbf{Within bins (High Correlation):} Since the treatment effect is homogeneous and the bin diameter is controlled, information is effectively shared. Identifying the optimal treatment for a single representative $X$ allows us to generalize the policy to all other $X'$ in the same bin with high confidence.\item \textbf{Between bins (Independence):} Separate bins can be treated as effectively independent. This decoupling ensures that the difficulty of estimation in regions with small gaps (where $\Delta(X)$ is small) does not contaminate the estimation in regions with large gaps, thereby balancing the bias-variance trade-off locally.\end{itemize}This delicate binning strategy, which optimally exploits the smoothness of the reward function, aligns with techniques in the non-parametric contextual bandit literature \citep{hu2020smooth, li2019dimension}.

We now detail the specific implementation of the \textit{ConSE} algorithm, formally presented in Algorithm \ref{alg:conse-phase1}, \ref{alg:conse-phase2-beta1}. 
The procedure employs a batched adaptive binning and elimination strategy. We define the initial batch length as $r = 2^{d+2}$, with the duration of the $m$-th batch scaling geometrically as $r^m$.
The experimental timeline is structured into two distinct phases:

\begin{enumerate}
    \item \textbf{Phase I: Regret Minimization via Adaptive Refinement.} 
    During the initial phase of the experiment consists of $O(\log n)$ batches, the algorithm iteratively refines the partition of the covariate space. At the conclusion of each batch, we evaluate every active bin. If the scale of the treatment effect within a bin is comparable to the bin diameter (ensuring the bias-variance trade-off is balanced) and the superior treatment can be identified with high probability, we terminate further splitting for this bin. The bin is then marked as ``inactive,'' and the algorithm commits to exploiting the optimal treatment in this region for the remainder of the phase. Conversely, if these conditions are not met, the bin is subdivided into smaller hypercubes to reduce bias, utilizing the subsequent batch to estimate the treatment effect at a finer resolution. This phase is critical for learning the intrinsic structural complexity of the instance and minimizing cumulative regret by rapidly pruning suboptimal arms in easy regions.
    This theoretical finding has profound implications for \textbf{algorithmic equity} in healthcare and policy-making. It suggests that ensuring fair representation in data collection does not necessarily require a sacrifice of aggregate welfare. By ``leveling up'' the exploration in easier subgroups (e.g., majority demographics) to match the learning requirements of harder groups, decision-makers prevent the systemic under-studying of specific populations. This ensures that the final deployed policy is robust across all demographics, not just the ``average'' user, thereby aligning statistical optimality with ethical imperatives at a negligible asymptotic cost.
    \item \textbf{Phase II: Estimation via Controlled Randomization (The RCT Phase).} 
    In the subsequent phase, leveraging the granular binning structure learned in Phase I, the focus shifts to precise CATE estimation. The experimenter specifies an exploration parameter $\alpha$, which reflects the desired balance between estimation accuracy and welfare preservation. Based on $\alpha$, the algorithm calibrates a specific randomization probability for each bin. This probability governs a localized Randomized Controlled Trial (RCT), ensuring that sufficient exploration is maintained to satisfy the estimation requirements without incurring excessive welfare loss.
    From a managerial perspective, this result exposes the hidden danger of \textbf{``short-termism''} in standard A/B testing tools that aggressively optimize for immediate conversion rates. While such algorithms minimize the operational cost of the experiment, they often fail to generate the high-fidelity data required to build a truly personalized product strategy. Our framework proves that a calculated increase in exploration cost—investing in learning even when the answer seems obvious—is a necessary investment to secure superior long-term deployment value. In essence, strictly minimizing regret is akin to minimizing R\&D costs at the expense of product quality.
\end{enumerate}


\begin{algorithm}[!t]
\caption{ConSE: Phase I (Batched Segmentation and Elimination)}
\label{alg:conse-phase1}
\begin{algorithmic}[1]
\REQUIRE Horizon $n$, dimension $d$.
\STATE Set $r \gets 2^{d+2}$ and let $H$ be the largest integer such that $\sum_{j=1}^{H} r^{j}\leqslant n/2$.
\STATE Initialize partition $\mathcal{P}\gets\{\mathcal{X}\}$; mark every $S\in\mathcal{P}$ as \textsc{NonTerminal}. Use the first $\sum_{j=1}^{H} r^{j}$ data and divide them to $H$ batches.
\FOR{Batch $i=1,2,\ldots,H$} \label{line:batch-loop}
    \FOR{each round in batch $i$}
        \STATE Observe covariate $x$; let $S\in\mathcal{P}$ be the unique cube with $x\in S$.
        \IF{$S$ is \textsc{Terminal}}
            \STATE Play the arm labeled \textsc{Optimal} for $S$.
        \ELSE
            \STATE Sample arm $A\in\{1,2\}$ uniformly (probabilities $(\tfrac12,\tfrac12)$) and play $A$.
        \ENDIF
        \STATE Observe reward.
    \ENDFOR
    \FOR{each $S\in\mathcal{P}$ that is \textsc{NonTerminal}}
        \STATE Average \emph{only} data from batch $i$ with covariates $x\in S$ and arm $j\in\{1,2\}$ to compute $\widehat{\mu}^{\,i}_j(S)$,\\
        \begin{center}
        $
        \widehat{\Delta}_i(S)\gets \widehat{\mu}^{\,i}_2(S)-\widehat{\mu}^{\,i}_1(S).
        $
        \end{center}
        \IF{$\widehat{\Delta}_i(S)>\dfrac{\log n}{2^{i}}$}
            \STATE Mark $S$ as \textsc{Terminal} and label arm $2$ as \textsc{Optimal}.
        \ELSIF{$\widehat{\Delta}_i(S)<-\dfrac{\log n}{2^{i}}$}
            \STATE Mark $S$ as \textsc{Terminal} and label arm $1$ as \textsc{Optimal}.
        \ELSE
            \STATE Subdivide $S$ into $2^{d}$ dyadic sub-cubes and replace $S$ in $\mathcal{P}$ by them
            (all marked \textsc{NonTerminal}).
        \ENDIF
    \ENDFOR
\ENDFOR
\STATE \textbf{Output:} partition $\mathcal{P}$ with terminal/optimal labels.
\end{algorithmic}
\end{algorithm}


\begin{algorithm}[!t]
\caption{ConSE: Phase II (Controlled Exploitation and Refined Estimation)}
\label{alg:conse-phase2-beta1}
\begin{algorithmic}[1]
\REQUIRE Horizon $n$, partition $\mathcal{P}$ and terminal/optimal labels from Phase I,
dimension $d$.
\STATE Suppose $\mathcal{P}$ includes $C_k$ level-$k$ cubes, define $\Delta_k=a_k=2^{-k}$, $p_k=C_ka_k^d$ ($1\leqslant k\leqslant H$). Calculate:\\
\begin{center}
    $\Gamma_1=n^\frac{1}{2+d}\left(\sum_{k<H} p_k\Delta_k^{\frac{2}{4+d}}\right)^\frac{4+d}{2},\ \Gamma_2=n^\frac{1}{2+d}\sum_{k<H} p_k\Delta_k^{-1-d}$
\end{center}
\STATE \textbf{Define:} $E_{\min}=n^{-\frac{2}{d+2}}$, $\hat{E}_{\max}=\left(\frac{\Gamma_1}{\Gamma_2\vee n \cdot p_H}\right)^{\frac{2}{d+2}}\vee n^{-\frac{2}{d+2}}$.
\STATE Report \textbf{YES} if $p_H\leqslant \frac{\Gamma_2}{n}$ or $p_H\geqslant \Gamma_1$, else report \textbf{NO}.
\STATE \textbf{Choose:} a desired error level $\tilde{E}\in[E_{\min},\hat{E}_{\max}]$.
\STATE Let $T_1 \gets \sum_{j=1}^{H} r^{j}$ and $T_2 \gets n-T_1$, where $r=2^{d+2}$.
\STATE For $k<H$, set
\[
q_k \;\gets\; \frac{1}{2n}\left(\frac{p_k}{\tilde{E}}\right)^{\frac{2+d}{2}},
\quad \text{and} \quad
q_H \gets \frac12 .
\]
\FOR{$t=T_1+1,\ldots,n$}
    \STATE Observe covariate $x$ and find the unique cube $S\in\mathcal{P}$ with $x\in S$;
    let $k$ be the level of $S$.
    \IF{$k<H$}
        \STATE Play the arm labeled \textsc{Optimal} on $S$ with probability $1-q_k$
        (and the other arm with probability $q_k$).
    \ELSE
        \STATE Sample an arm uniformly from $\{1,2\}$ and play it.
    \ENDIF
    \STATE Observe reward.
\ENDFOR
\FOR{each level $k\in\{0,1,\ldots,H\}$}
    \STATE Set $M_k \;\gets\; \lfloor a_k\,(n q_k)^{\frac{1}{2+d}}\rfloor$.
    \STATE Subdivide every level-$k$ cube $S\in\mathcal{P}$ into $M_k^{d}$ dyadic sub-cubes.
    \STATE Independently estimate the mean reward of each arm within every resulting sub-cube
    using Phase II samples falling in that sub-cube.
\ENDFOR
\STATE \textbf{Output:} refined partition and local mean estimates.
\end{algorithmic}
\end{algorithm}

In the following theorem, we show that for a designed estimation error belongs to a range $[E_{\min},\hat{E}_{\max}]$, where the two endpoints represent the most loss-sensitive and the most accuracy-sensitive extremes, \textit{ConSE} achieves the optimal tradeoff as characterized in Theorem \ref{thm-lower}, thus achieving the whole achievable Pareto optimal frontier.

\begin{theorem}\label{thm_upper_noprivacy}
Fix an instance $\nu$, and any target accuracy parameter $\tilde{E}\geqslant E_{\min}$, the proposed algorithm guarantees,
\[
R_{n,\nu}
=
\softO\!\left(
\tilde{E}^{-\frac{2+d}{2}}\,H(\nu)
\;\vee\;
R_{\min}
\right),
\qquad
E_{n,\nu}
=
\softO\!\left(
\tilde{E}
\right).
\]
Moreover, when $\tilde{E}\in [E_{\min},E_{\max}]$, the resulting trade-off pair $(R,E)$ is Pareto optimal.
Here,
\[
E_{\max}
=
R_{\min}^{-\frac{2}{2+d}}(\nu)\,H(\nu)^{\frac{2}{2+d}}
\;\vee\;
E_{\min}
\]
is an instance-dependent upper limit on the achievable estimation accuracy and is generally unknown.
\end{theorem}

Despite the analytical complexity arising from the intricate nature of adaptive experimentation, our theoretical results yield compelling insights that offer practical guidance for policy-making.

\noindent \textbf{1. Implicit Group Fairness in Pareto-Optimal Designs.} While we do not explicitly enforce fairness constraints, our analysis reveals that \textbf{any Pareto-optimal experiment naturally tends toward group fairness in regret}. Since our framework partitions the feature space into $O(\log n)$ distinct bin types, consider a scenario where the cumulative regret is concentrated heavily in one specific group while remaining negligible in others. In such a case, increasing the exploration budget for the ``low-regret'' groups to match the ``high-regret'' group would not increase the order of the total regret (which is dominated by the maximum term), yet it would significantly reduce the estimation error within those under-explored groups. Consequently, the original allocation is not Pareto optimal. Operationally, this implies a principle of \emph{fairness in welfare loss}: to achieve optimality on the Pareto frontier, the decision-maker should allocate risk (regret) broadly across subgroups, rather than penalizing difficult-to-learn groups while completely neglecting exploration in easier populations.

\noindent \textbf{2. Neither Pure Regret Minimization or RCT is  Pareto Optimal.} Counter-intuitively, the extremes of the Pareto frontier are not simply defined by pure regret minimization algorithms (e.g., \citealt{hu2020smooth, simchi2022bypassing}) versus completely randomized trials. In fact, \textbf{standard regret minimization algorithms are strictly suboptimal} in the dual-objective landscape for most instances. Consider a heterogeneous instance where half of the feature space exhibits a constant treatment effect (``easy'' region, $\Delta(X) \approx 1$), while the other half corresponds to the hardest learning regime with a gap scaling as $\Delta(X) \approx n^{-\frac{1}{d+2}}$. In this scenario, a standard regret minimizer incurs negligible regret $\widetilde{O}(1)$ (or logarithmic) on the easy half, but suffers a large minimax regret of $\widetilde{O}(n^{\frac{d+1}{d+2}})$ on the hard half. The total regret is thus dominated by the hard region: $\widetilde{O}(n^{\frac{d+1}{d+2}})$. However, because the algorithm ruthlessly exploits the easy region, it collects almost no samples from the suboptimal arm, resulting in a large estimation error (approaching $\widetilde{O}(1)$). Crucially, if we were to increase exploration in the easy region to also incur $\widetilde{O}(n^{\frac{d+1}{d+2}})$ regret, the total network regret would remain $\widetilde{O}(n^{\frac{d+1}{d+2}})$ (order-wise unchanged), but the estimation error in the easy region would drop drastically from $\widetilde{O}(1)$ to $\widetilde{O}(n^{-\frac{2(d+1)}{(d+2)^2}})$. Similarly one can construct a case where complete A/B testing is not Pareto optimal. This demonstrates that the optimal strategy involves a distinct explore-exploit trade-off when optimizing dual objectives, requiring sufficient exploration even in regions where the optimal action is easily identified. 

\begin{figure}[htbp]
  \centering
  \begin{minipage}[b]{0.45\textwidth}
    \centering
    \includegraphics[width=\linewidth]{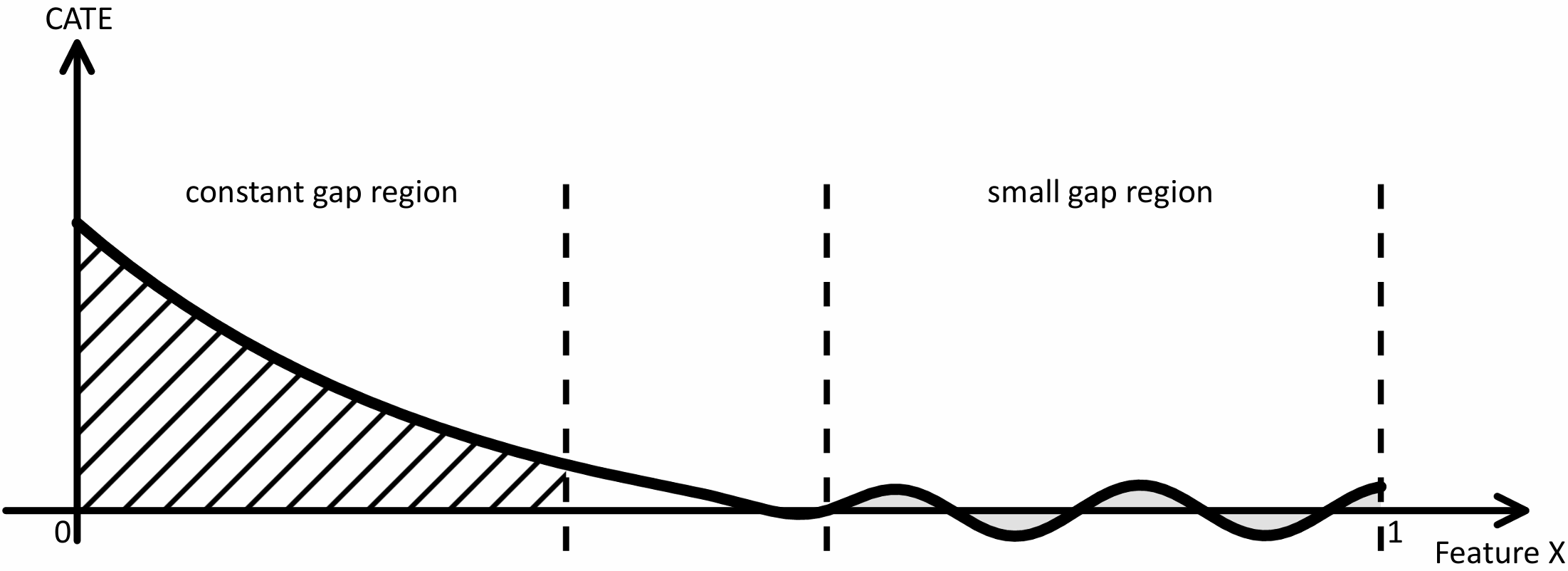}
    \caption{Instance with small gap}
    \label{fig:smallgap}
  \end{minipage}
  \hfill
  \begin{minipage}[b]{0.45\textwidth}
    \centering
    \includegraphics[width=\linewidth]{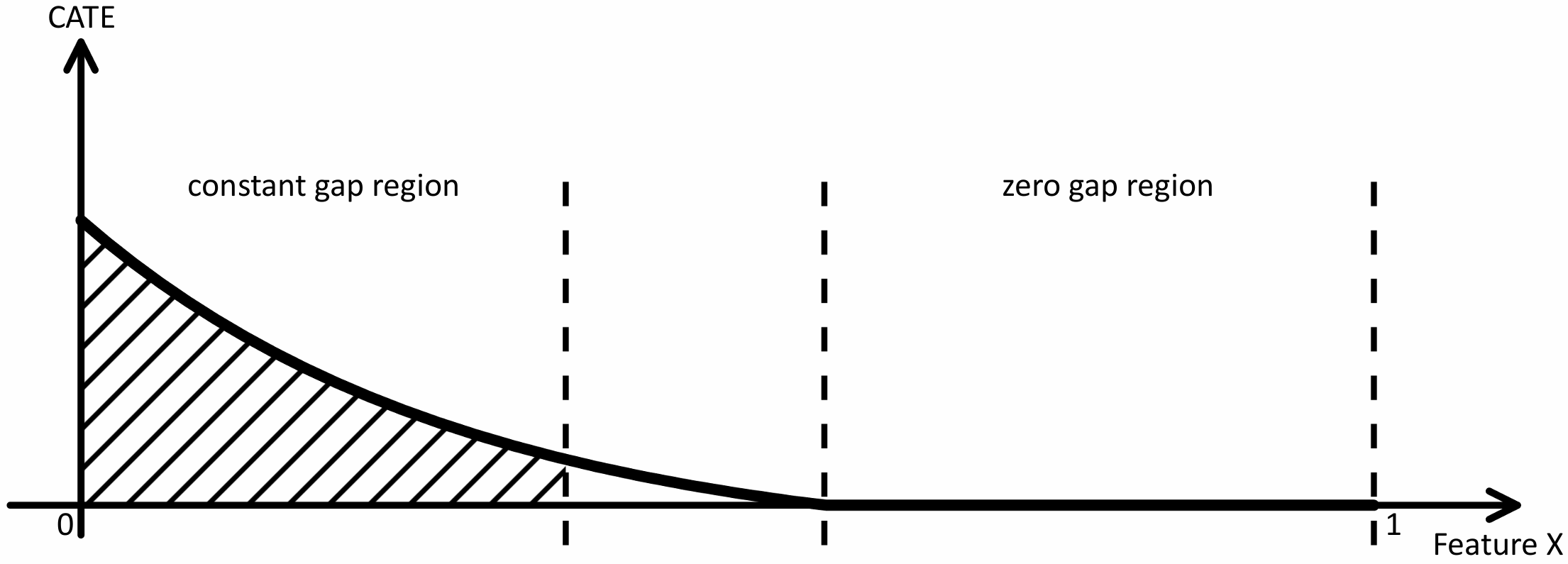}
    \caption{Instance with zero gap}
    \label{fig:zerogap}
  \end{minipage}
\end{figure}

\noindent\textbf{3. Not the whole Pareto optimal curve is achievable or even learnable.}
While Theorem 2 characterizes the full Pareto frontier, we must address the reality that this curve is not always fully learnable or achievable due to the inherent limits of statistical resolution. A central challenge in adaptive experimentation is that the algorithm may encounter a family of instances $\nu'$ that are statistically indistinguishable from the true instance $\nu$ given $n$ samples. Consider again the example where half of feature spaces has constant treatment effect, while the other half effect $\widetilde{O}(n^{-\frac{1}{d+2}})$. Note that in a $d$-dimensional covariate space with Lipschitz rewards, a treatment effect of magnitude $n^{-\frac{1}{d+2}}$ is often indistinguishable from a zero effect within the noise floor. However, these indistinguishable cases can possess drastically different Pareto frontiers (See Figure \ref{fig:multiple_pareto} for an illustration of different instances with different Pareto Frontiers that are not distinguishable):
\begin{figure}
    \centering
    \includegraphics[width=0.6\linewidth]{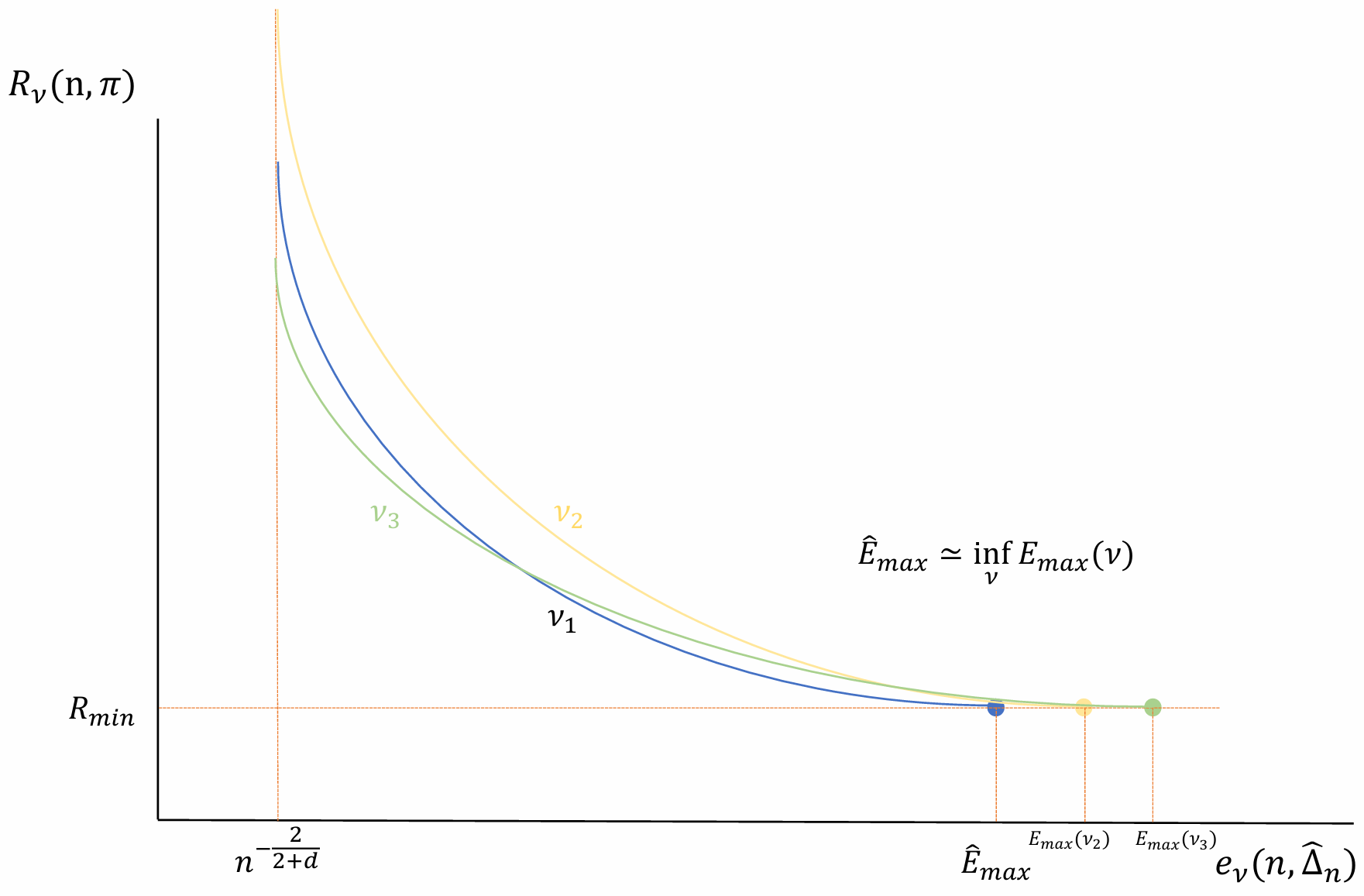}
    \caption{Multiple Indistinguishable Pareto Frontiers}
    \label{fig:multiple_pareto}
\end{figure}
\begin{enumerate}
    \item The ``Gap'' Case: If the true effect is $n^{-\frac{1}{d+2}}$, this is the \emph{worst} case in regret minimization, and a $\widetilde{O}(n^{d+1/d+2})$ regret is unavoidable. Therefore, for the half feature space with constant treatment effect, we need to increase exploration to also incur $\widetilde{O}(n^{d+1/d+2})$ regret, as explained above why \textbf{regret minimization is strictly dominated}.
    \item The ``Null'' Case: If the effect is $0$, the cumulative regret is inherently $0$ regardless of the allocation for the small gap region, and \textbf{it is Pareto optimal to simply do regret minimization} in the constant effect region, since now the regret only comes from the constant gap region. However, these two cases can not be distinguished by any possible algorithm (see Fig \ref{fig:smallgap}, \ref{fig:zerogap}).
\end{enumerate}
Because the algorithm cannot definitively pick between these scenarios, it must find a range of choice parameters that \textbf{remain optimal across all possible indistinguishable alternatives}. To navigate this uncertainty, we define the equivalence class $\mathcal{P}(\nu)$ as the set of all instances sharing the same adaptive partition (as per Definition~\ref{def: Adaptive Partition}), representing the ``blind spot'' of statistical indistinguishability for $\nu$. Our algorithm produces a data-driven estimator $\hat{E}_{\max}$ that serves as a robust upper bound for the achievable error. We prove that with probability at least $1-\frac{1}{n^2}$, this estimator satisfies: $$\hat{E}_{\max} \;\asymp\; \inf_{\nu' \in \mathcal{P}(\nu)} E_{\max}(\nu').$$ 
In this framework, $\hat{E}_{\max}$ acts as an optimal conservative estimator. By selecting a target error level $\tilde{E} \in [E_{\min}, \hat{E}_{\max}]$, the algorithm attains a Pareto-optimal trade-off between regret $R$ and estimation error $E$ without requiring oracle knowledge of the instance's complexity. The algorithm explicitly tracks the possibility of multiple indistinguishable but divergent cases. If the algorithm outputs YES, it indicates that $E_{\max}(\nu')$ is identical for all $\nu' \in \mathcal{P}(\nu)$, which implies that our estimator $\hat{E}_{\max}$ has converged to the true $E_{\max}$. In such scenarios, the algorithm successfully recovers the entire Pareto-optimal frontier. The use of the $\inf$ over the equivalence class ensures that the decision-maker is protected against the most ``pessimistic'' indistinguishable instance. This provides a robustness guarantee: you are not just optimal for the instance you think you see, but for all instances that are statistically consistent with your observations. For practitioners, this resolves the ``over-fitting'' risk in experimental design. Instead of aggressively optimizing for a singular point estimate of the treatment effect, the algorithm identifies a stable operating zone from $E_{\min}$ to $\hat{E}_{\max}$ where the trade-off remains efficient even under statistical uncertainty.

\section{DP-ConSE: An Optimal Adaptive Experimentation Framework in Privacy Protection and Post-experiment Performance }\label{sec:privacy_simple_regret}
In this section, we address the final two questions posed in Section \ref{sec:intro}: quantifying the cost of privacy protection, and determining whether the dual constraints of in-experiment welfare loss and privacy compromise the post-experiment welfare of the broader population. Our analysis yields highly favorable results on both fronts. We introduce \textit{DP-ConSE} (Algorithm \ref{alg:dp-conse-phase1}, \ref{alg:dp-conse-phase2}), a privacy-preserving adaptive experimentation framework that achieves differential privacy essentially ``for free.'' Specifically, we prove that the additional deterioration in estimation accuracy and cumulative regret attributable to privacy constraints is asymptotically negligible. Furthermore, we demonstrate that the policy derived from the privately collected data of \textit{DP-ConSE} attains the optimal possible post-experiment welfare. This implies a fundamental alignment of objectives: when optimizing the delicate trade-off between experimental regret and estimation accuracy, decision-makers need not fear a degradation in long-term deployment utility or incur a prohibitive cost for privacy compliance.


\begin{algorithm}[!t]
\caption{DP-ConSE: Phase I (Batched Segmentation and Elimination)}
\label{alg:dp-conse-phase1}
\begin{algorithmic}[1]
\REQUIRE Horizon $n$, dimension $d$, privacy parameter $\varepsilon>0$.
\STATE Set $r \gets 2^{d+2}$ and let $H$ be the largest integer such that $\sum_{j=1}^{H} r^{j}\leqslant n/2$.
\STATE Initialize partition $\mathcal{P}\gets\{\mathcal{X}\}$; mark every $S\in\mathcal{P}$ as \textsc{NonTerminal}. Use the first $\sum_{j=1}^{H} r^{j}$ data and divide them to $H$ batches.
\FOR{Batch $i=1,2,\ldots,H$} \label{line:dp-batch-loop}
    \FOR{each round in batch $i$}
        \STATE Observe covariate $x$; let $S\in\mathcal{P}$ be the unique cube with $x\in S$.
        \IF{$S$ is \textsc{Terminal}}
            \STATE Play the arm labeled \textsc{Optimal} for $S$.
        \ELSE
            \STATE Sample each arm $A\in\{1,2\}$ uniformly (half-half in the whole batch) and play $A$.
        \ENDIF
        \STATE Observe reward.
    \ENDFOR
    \FOR{each $S\in\mathcal{P}$ that is \textsc{NonTerminal}}
        \STATE Average \emph{only} data from batch $i$ with covariates $x\in S$ and arm $j\in\{1,2\}$ to compute $\widehat{\mu}^{\,i}_j(S)$,\\
        \begin{center}
        $
        \widehat{\Delta}_i(S)\gets \widehat{\mu}^{\,i}_2(S)-\widehat{\mu}^{\,i}_1(S)
        + \mathrm{Lap}\!\left(\dfrac{2}{r^{i}\varepsilon}\right).
        $
        \end{center}
        \IF{$\widehat{\Delta}_i(S)>\dfrac{(\log n)^2}{2^{i}}$}
            \STATE Mark $S$ as \textsc{Terminal} and label arm $2$ as \textsc{Optimal}.
        \ELSIF{$\widehat{\Delta}_i(S)<-\dfrac{(\log n)^2}{2^{i}}$}
            \STATE Mark $S$ as \textsc{Terminal} and label arm $1$ as \textsc{Optimal}.
        \ELSE
            \STATE Subdivide $S$ into $2^{d}$ dyadic sub-cubes and replace $S$ in $\mathcal{P}$ by them
            (all marked \textsc{NonTerminal}).
        \ENDIF
    \ENDFOR
\ENDFOR
\STATE \textbf{Output:} partition $\mathcal{P}$ with terminal/optimal labels.
\end{algorithmic}
\end{algorithm}


\begin{algorithm}[!t]
\caption{DP-ConSE: Phase II (Controlled Exploitation and Private Refined Estimation)}
\label{alg:dp-conse-phase2}
\begin{algorithmic}[1]
\REQUIRE Horizon $n$, dimension $d$, privacy parameter $\varepsilon>0$, and partition $\mathcal{P}$ from Phase I.
\STATE Suppose $\mathcal{P}$ includes $C_k$ level-$k$ cubes, define $\Delta_k=a_k=2^{-k}$, $p_k=C_ka_k^d$ ($1\leqslant k\leqslant p$). Calculate:\\
\begin{center}
    $\Gamma_1=n^\frac{1}{2+d}\left(\sum_{k<H} p_k\Delta_k^{\frac{2}{4+d}}\right)^\frac{4+d}{2},\ \Gamma_2=n^\frac{1}{2+d}\sum_{k<H} p_k\Delta_k^{-1-d}$
\end{center}
\STATE \textbf{Define:} $E_{\min}=n^{-\frac{2}{d+2}}$, $\hat{E}_{\max}=\left(\frac{\Gamma_1}{\Gamma_2\vee n \cdot p_H}\right)^{\frac{2}{d+2}}\vee n^{-\frac{2}{d+2}}$.
\STATE Report \textbf{YES} if $p_H\leqslant \frac{\Gamma_2}{n}$ or $p_H\geqslant \Gamma_1$, else report \textbf{NO}.
\STATE \textbf{Choose:} a desired error level $\tilde{E}\in[E_{\min},\hat{E}_{\max}]$.
\STATE Let $T_1 \gets \sum_{j=1}^{H} r^{j}$ and $T_2 \gets n-T_1$, where $r=2^{d+2}$.
\STATE For $k<p$, set
\[
q_k \;\gets\; \frac{1}{2n}\left(\frac{p_k}{\tilde{E}}\right)^{\frac{2+d}{2}},
\quad \text{and} \quad
q_H \gets \frac12 .
\]
\FOR{$t=T_1+1,\ldots,n$}
    \STATE Observe covariate $x$; let $S\in\mathcal{P}$ be the unique cube with $x\in S$, and let $k$ be its level.
    \IF{$k<H$}
        \STATE Play the arm labeled \textsc{Optimal} for $S$ with probability $1-q_k$.
    \ELSE
        \STATE Sample an arm uniformly from $A\in\{1,2\}$ and play it.
    \ENDIF
    \STATE Observe reward.
\ENDFOR

\FOR{level $k=0,1,\ldots,H$}
    \STATE Set $M_k\;\gets\;\lfloor a_k\,(n q_k)^{\frac{1}{2+d}}\rfloor$.
    \STATE Subdivide each level-$k$ cube in $\mathcal{P}$ into $M_k^{d}$ dyadic sub-cubes.
    \FOR{each resulting sub-cube $B$}
        \FOR{arm $j\in\{1,2\}$}
            \STATE Let $s_j(B)$ and $c_j(B)$ be the Phase II reward sum and pull count for arm $j$ within $B$.
            \STATE Output the private mean estimate $\widetilde{\mu}_j(B)\;\gets\;
            \frac{s_j(B)+\mathrm{Lap}(2/\varepsilon)}
                 {\big(c_j(B)+\mathrm{Lap}(2/\varepsilon)\big)\vee 1}$.
        \ENDFOR
    \ENDFOR
\ENDFOR
\STATE \textbf{Output:} refined partition and private local mean estimates $\{\widetilde{\mu}_j(B)\}$.
\end{algorithmic}
\end{algorithm}

Crucially, the output of our experiment comprises both the adaptive allocation trajectory and the final estimation of the CATE. Consequently, the private information inherent in the covariate $X$ and outcome $Y$ must be shielded from leakage through either channel. Regarding the allocation mechanism, we privatize the ``sufficient statistics'' employed in adaptive binning—specifically, the estimated average treatment effect within each bin. Intuitively, we inject calibrated noise into these statistics to ensure that if the private data of any individual $(X_i, Y_i)$ were replaced by another $(X_{i'}, Y_{i'})$, an adversarial attacker would be statistically unable to distinguish whether the observed variation stems from the data change or the added noise. This guarantees that individual participation remains protected against inference attacks. Moreover, our batched design strategically leverages the fact that as the sample size within a batch increases, the sensitivity of the sufficient statistic to any single individual decreases. Consequently, the magnitude of the required noise scales down, becoming an asymptotically negligible lower-order term compared to the non-private statistical estimation error. Thus, the additional regret incurred by privacy preservation is minimal.The exponential growth of batch lengths plays a pivotal role in optimizing this trade-off between differential privacy protection and regret minimization. This design philosophy resonates with the DP Successive Elimination'' algorithm \citep{sajed2019optimal} and the widely adopted tree mechanism'' \citep{chan2011private} in DP-bandit literature. Similarly, we inject noise into the final CATE estimation in \textit{ConSE} to ensure end-to-end privacy, again ensuring the noise scale remains subordinate to the estimation error. We formalize these guarantees in the following theorem, demonstrating that \textit{DP-ConSE} achieves the same theoretical bounds as \textbf{ConSE}, effectively providing privacy protection at no additional cost to the convergence rate.
\begin{theorem}\label{thm_upper_privacy}
Fix an instance $\nu$, any privacy parameter $\varepsilon>0$ and any target accuracy parameter $\tilde{E}\geqslant E_{\min}$, the proposed algorithm is $\varepsilon$-DP and guarantees,
\[
R_{n,\nu}
=
\softO\!\left(
\tilde{E}^{-\frac{2+d}{2}}\,H(\nu)
\;\vee\;
R_{\min}
\right),
\qquad
E_{n,\nu}
=
\softO\!\left(
\tilde{E}
\right).
\]
Moreover, when $\tilde{E}\in [E_{\min},E_{\max}]$, the resulting trade-off pair $(R,E)$ is Pareto optimal.
Here,
\[
E_{\max}
=
R_{\min}^{-\frac{2}{2+d}}(\nu)\,H(\nu)^{\frac{2}{2+d}}
\;\vee\;
E_{\min}
\]
is an instance-dependent upper limit on the achievable estimation accuracy and is generally unknown.

Let $\hat{E}_{\max}$ denote the data-driven estimator of $E_{\max}$ produced by the algorithm.
Then, with probability of at least $1-\frac{1}{n}$,
\[
\hat{E}_{\max}
\;\asymp\;
\inf_{\nu'\in \mathcal{P}(\nu)} E_{\max}(\nu'),
\]
where $\mathcal{P}(\nu)$ is the equivalence class of instances sharing the same adaptive partition
(Definition~\ref{def: Adaptive Partition}), representing all statistically indistinguishable alternatives to $\nu$.
In this sense, $\hat{E}_{\max}$ is an optimal conservative estimator of $E_{\max}$.

Consequently, by selecting $\tilde{E}\in [E_{\min},\hat{E}_{\max}]$, the algorithm attains Pareto-optimal
$(R,E)$ trade-offs without prior knowledge of $E_{\max}$.
Furthermore, if the algorithm outputs \textbf{YES}, then $E_{\max}(\nu')$ is identical for all
$\nu'\in\mathcal{P}(\nu)$, implying $\hat{E}_{\max}\asymp E_{\max}$.
In this case, the algorithm recovers the entire Pareto-optimal frontier.
\end{theorem}

Ultimately, the overarching objective of most A/B testing is to identify a personalized treatment policy that maximizes long-term reward, such as revenue or therapeutic outcomes. Consequently, it is imperative to assess whether optimizing for the interim objectives discussed thus far—CATE estimation accuracy, cumulative regret, and privacy preservation—comes at the detriment of post-experiment welfare.
The following theorem establishes that our privacy-preserving, regret-optimal framework, \textit{DP-ConSE}, attains the theoretically optimal post-experiment welfare loss, \emph{irrespective} of the chosen exploration hyper-parameter $\alpha$. This result underscores a fundamental insight: the trade-off between estimation accuracy and regret is effectively an \textbf{intra-experimental} calibration. It serves to balance costs during the learning phase without negatively propagating to the \textbf{post-experiment} deployment efficiency.
\begin{theorem}\label{thm:simple_regret}
    For any $X\in\X$, the simple regret of choosing the “optimal” arm is at most 
    $$
    \softO\left(\frac{1}{n}+\E_X\left[|\Delta(X)|1\{|\Delta(X)|\leqslant 2(\log n)^2 n^{-\frac{1}{2+d}}\}\right]\right).
    $$
\end{theorem}
This result matches the existing result in \cite{simple-regret-instance-lowerbound} of $O(n^{-\frac{1+\alpha}{2+d}})$ for an instance $\nu$ with margin parameter $\alpha$, thus is instance-optimal.
 A pivotal observation underlying the proof of Theorem \ref{thm:simple_regret} is that minimizing the expected loss of a learned policy is typically achieved via two distinct classical paradigms:
the first is \emph{pure exploration} (e.g., uniform sampling), which aims to identify the optimal treatment with high probability;
the second is \emph{regret minimization}, which leverages online-to-batch conversion \citep{foster2020beyond} to output the average policy over the time horizon.
Our adaptive experimentation framework can be interpreted as a hybrid of these two extremes. Crucially, the logic follows that at least one of these modalities—either significant exploration or regret-minimizing exploitation—must be dominant for a substantial fraction of the experiment (i.e., at least $n/2$ rounds). Since both approaches independently yield optimal convergence rates for post-experiment policy loss (simple regret), their mixture guarantees rate-optimal performance.
This independence result provides a crucial \textbf{``safety net''} for practitioners. It resolves the anxiety associated with hyper-parameter tuning in adaptive systems: whether a manager prioritizes user safety (high $\alpha$) or data granularity (low $\alpha$) during the trial based on immediate operational constraints, the resulting policy is guaranteed to converge to the same optimal performance upon full launch. This allows organizations to flexibly align experimental design with their budget or safety protocols without fearing that they are effectively compromising the long-term quality of the AI product.

\section{Conclusion}\label{sec: conclusion}
This paper addresses the complexity of optimizing adaptive experiments under the competing constraints of welfare maximization, causal inference precision, and privacy protection. By analyzing the problem through a non-parametric lens, we established the information-theoretic lower bounds that govern the trade-off between cumulative regret and CATE estimation error. Our proposed framework, ConSE, not only achieves this Pareto optimality through adaptive covariate segmentation but also seamlessly integrates differential privacy via DP-ConSE without compromising convergence rates.

A pivotal finding of our work is the decoupling of experimental objectives from post-experimental outcomes. We show that the trade-off between regret and estimation is an intra-experimental calibration; as long as the design lies on the Pareto frontier, the resulting policy is guaranteed to be optimal for full-scale deployment. This offers a compelling practical guideline for policymakers in digital platforms and clinical research: experiments can be tuned to minimize immediate participant harm or maximize privacy protection without fear of degrading the long-term efficacy of the deployed intervention. Future work may extend this framework to settings with network interference or non-stationary environments.

\bibliographystyle{plainnat}
\bibliography{citation}

@article{dietvorst2015algorithm,
  author  = {Berkeley J. Dietvorst and Joseph P. Simmons and Cade Massey},
  title   = {Algorithm Aversion: People Erroneously Avoid Algorithms After Seeing Them Err},
  journal = {Journal of Experimental Psychology: General},
  year    = {2015},
  volume  = {144},
  number  = {1},
  pages   = {114--126},
  doi     = {10.1037/xge0000033}
}

@article{dwork2006calibrating,
  title={Calibrating noise to sensitivity in private data analysis},
  author={Dwork, Cynthia and McSherry, Frank and Nissim, Kobbi and Smith, Adam},
  booktitle={Theory of cryptography conference},
  pages={265--284},
  year={2006},
  organization={Springer}
}

@inproceedings{erlingsson2014rappor,
  title={Rappor: Randomized aggregatable privacy-preserving ordinal response},
  author={Erlingsson, {\'U}lfar and Pihur, Vasyl and Korolova, Aleksandra},
  booktitle={Proceedings of the 2014 ACM SIGSAC conference on computer and communications security},
  pages={1054--1067},
  year={2014}
}

@article{homer2008resolving,
  title={Resolving individuals contributing trace amounts of DNA to highly complex mixtures using high-density SNP genotyping microarrays},
  author={Homer, Nils and Szelinger, Szabolcs and Redman, Margot and Duggan, David and Tembe, Waibhav and Muehling, Jill and Pearson, William and Stephan, Dietrich and Nelson, Stanley and Craig, David},
  journal={PLoS genetics},
  volume={4},
  number={8},
  pages={e1000167},
  year={2008}
}

@article{narayanan2008robust,
  title={Robust de-anonymization of large sparse datasets},
  author={Narayanan, Arvind and Shmatikov, Vitaly},
  booktitle={2008 IEEE Symposium on Security and Privacy},
  pages={111--125},
  year={2008},
  organization={IEEE}
}

@book{lattimore2020bandit,
  title={Bandit algorithms},
  author={Lattimore, Tor and Szepesv{\'a}ri, Csaba},
  year={2020},
  publisher={Cambridge University Press}
}

@article{auer2002finite,
  title={Finite-time analysis of the multiarmed bandit problem},
  author={Auer, Peter and Cesa-Bianchi, Nicolo and Fischer, Paul},
  journal={Machine learning},
  volume={47},
  pages={235--256},
  year={2002}
}

@article{streitfeld2000amazon,
  title={On the Web, Price Tags Blur; Amazon.com Test Varies Costs for Same Items},
  author={Streitfeld, David},
  journal={The Washington Post},
  volume={27},
  year={2000},
  note={Retrieved from strict regulatory and public archives regarding the Amazon DVD pricing controversy}
}

@article{li2024optimal,
  title={Optimal adaptive experimental design for estimating treatment effect},
  author={Li, Jiachun and Simchi-Levi, David and Zhao, Yunxiao},
  journal={arXiv preprint arXiv:2410.05552},
  year={2024}
}

@inproceedings{abbasi2011improved,
  title={Improved algorithms for linear stochastic bandits},
  author={Abbasi-Yadkori, Yasin and P{\'a}l, D{\'a}vid and Szepesv{\'a}ri, Csaba},
  booktitle={Advances in Neural Information Processing Systems},
  volume={24},
  pages={2312--2320},
  year={2011}
}

@inproceedings{lattimore2017end,
  title={The end of optimism? an asymptotic analysis of finite-armed linear bandits},
  author={Lattimore, Tor and Szepesvari, Csaba},
  booktitle={Artificial Intelligence and Statistics},
  pages={728--737},
  year={2017},
  organization={PMLR}
}

@inproceedings{hao2020adaptive,
  title={Adaptive exploration in linear contextual bandit},
  author={Hao, Botao and Lattimore, Tor and Szepesvari, Csaba},
  booktitle={International Conference on Artificial Intelligence and Statistics},
  pages={3536--3545},
  year={2020},
  organization={PMLR}
}

@inproceedings{chu2011contextual,
  title={Contextual bandits with linear payoff functions},
  author={Chu, Wei and Li, Lihong and Reyzin, Lev and Schapire, Robert},
  booktitle={Proceedings of the fourteenth international conference on artificial intelligence and statistics},
  pages={208--214},
  year={2011},
  organization={JMLR Workshop and Conference Proceedings}
}

@inproceedings{li2017provably,
  title={Provably optimal algorithms for generalized linear contextual bandits},
  author={Li, Lihong and Lu, Yu and Zhou, Dengyong},
  booktitle={International Conference on Machine Learning},
  pages={2071--2080},
  year={2017},
  organization={PMLR}
}

@inproceedings{foster2020beyond,
  title={Beyond ucb: Optimal and efficient contextual bandits with regression oracles},
  author={Foster, Dylan and Rakhlin, Alexander},
  booktitle={International conference on machine learning},
  pages={3199--3210},
  year={2020},
  organization={PMLR}
}

@article{simchi2022bypassing,
  title={Bypassing the monster: A faster and simpler optimal algorithm for contextual bandits under realizability},
  author={Simchi-Levi, David and Xu, Yunzong},
  journal={Mathematics of Operations Research},
  volume={47},
  number={3},
  pages={1904--1931},
  year={2022},
  publisher={INFORMS}
}

@article{rigollet2010nonparametric,
  title={Nonparametric bandits with covariates},
  author={Rigollet, Philippe and Zeevi, Assaf},
  journal={arXiv preprint arXiv:1003.1630},
  year={2010}
}

@article{perchet2013multi,
  title={The multi-armed bandit problem with covariates},
  author={Perchet, Vianney and Rigollet, Philippe},
  year={2013}
}

@inproceedings{hu2020smooth,
  title={Smooth contextual bandits: Bridging the parametric and non-differentiable regret regimes},
  author={Hu, Yichun and Kallus, Nathan and Mao, Xiaojie},
  booktitle={Conference on Learning Theory},
  pages={2007--2010},
  year={2020},
  organization={PMLR}
}

@article{li2019dimension,
  title={A dimension-free algorithm for contextual continuum-armed bandits},
  author={Li, Wenhao and Chen, Ningyuan and Hong, L Jeff},
  journal={arXiv preprint arXiv:1907.06550},
  year={2019}
}

@article{zuo2025pareto,
  title={On Pareto Optimality for the Multinomial Logistic Bandit},
  author={Zuo, Jierui and Qin, Hanzhang},
  journal={arXiv preprint arXiv:2501.19277},
  year={2025}
}

@article{duan2024regret,
  title={Regret minimization and statistical inference in online decision making with high-dimensional covariates},
  author={Duan, Congyuan and Ma, Wanteng and Jiang, Jiashuo and Xia, Dong},
  journal={arXiv preprint arXiv:2411.06329},
  year={2024}
}

@article{zhang2024online,
  title={Online experimental design with estimation-regret trade-off under network interference},
  author={Zhang, Zhiheng and Wang, Zichen},
  journal={arXiv preprint arXiv:2412.03727},
  year={2024}
}

@article{zhong2021achieving,
  title={Achieving the pareto frontier of regret minimization and best arm identification in multi-armed bandits},
  author={Zhong, Zixin and Cheung, Wang Chi and Tan, Vincent YF},
  journal={arXiv preprint arXiv:2110.08627},
  year={2021}
}

@article{zhao2023adaptive,
  title={Adaptive Neyman Allocation},
  author={Zhao, Jinglong},
  year={2023}
}

@misc{wager2024causal,
  title={Causal inference: A statistical learning approach},
  author={Wager, Stefan},
  year={2024},
  publisher={Technical report, Stanford University}
}

@article{dai2023clip,
  title={Clip-OGD: An Experimental Design for Adaptive Neyman Allocation in Sequential Experiments},
  author={Dai, Jessica and Gradu, Paula and Harshaw, Christopher},
  journal={arXiv preprint arXiv:2305.17187},
  year={2023}
}

@inproceedings{simchi2023multi,
  title={Multi-armed bandit experimental design: Online decision-making and adaptive inference},
  author={Simchi-Levi, David and Wang, Chonghuan},
  booktitle={International Conference on Artificial Intelligence and Statistics},
  pages={3086--3097},
  year={2023},
  organization={PMLR}
}

@article{shariff2018differentially,
  title={Differentially private contextual linear bandits},
  author={Shariff, Roshan and Sheffet, Or},
  journal={Advances in Neural Information Processing Systems},
  volume={31},
  year={2018}
}

@article{chan2011private,
  title={Private and continual release of statistics},
  author={Chan, T-H Hubert and Shi, Elaine and Song, Dawn},
  journal={ACM Transactions on Information and System Security (TISSEC)},
  volume={14},
  number={3},
  pages={1--24},
  year={2011},
  publisher={ACM New York, NY, USA}
}

@inproceedings{tossou2016algorithms,
  title={Algorithms for differentially private multi-armed bandits},
  author={Tossou, Aristide and Dimitrakakis, Christos},
  booktitle={Proceedings of the AAAI Conference on Artificial Intelligence},
  volume={30},
  number={1},
  year={2016}
}

@article{azize2022privacy,
  title={When privacy meets partial information: A refined analysis of differentially private bandits},
  author={Azize, Achraf and Basu, Debabrota},
  journal={Advances in Neural Information Processing Systems},
  volume={35},
  pages={32199--32210},
  year={2022}
}

@inproceedings{sajed2019optimal,
  title={An optimal private stochastic-mab algorithm based on optimal private stopping rule},
  author={Sajed, Touqir and Sheffet, Or},
  booktitle={International Conference on Machine Learning},
  pages={5579--5588},
  year={2019},
  organization={PMLR}
}

@article{hanna2022differentially,
  title={Differentially private stochastic linear bandits:(almost) for free},
  author={Hanna, Osama A and Girgis, Antonious M and Fragouli, Christina and Diggavi, Suhas},
  journal={arXiv preprint arXiv:2207.03445},
  year={2022}
}

@article{zheng2020locally,
  title={Locally differentially private (contextual) bandits learning},
  author={Zheng, Kai and Cai, Tianle and Huang, Weiran and Li, Zhenguo and Wang, Liwei},
  journal={Advances in Neural Information Processing Systems},
  volume={33},
  pages={12300--12310},
  year={2020}
}

@article{chen2022privacy,
  title={Privacy-preserving dynamic personalized pricing with demand learning},
  author={Chen, Xi and Simchi-Levi, David and Wang, Yining},
  journal={Management Science},
  volume={68},
  number={7},
  pages={4878--4898},
  year={2022},
  publisher={INFORMS}
}

@article{johari2015always,
  title={Always valid inference: Bringing sequential analysis to A/B testing},
  author={Johari, Ramesh and Pekelis, Leo and Walsh, David J},
  journal={arXiv preprint arXiv:1512.04922},
  year={2015}
}

@article{bojinov2021panel,
  title={Panel experiments and dynamic causal effects: A finite population perspective},
  author={Bojinov, Iavor and Rambachan, Ashesh and Shephard, Neil},
  journal={Quantitative Economics},
  volume={12},
  number={4},
  pages={1171--1196},
  year={2021},
  publisher={Wiley Online Library}
}

@article{bojinov2023design,
  title={Design and analysis of switchback experiments},
  author={Bojinov, Iavor and Simchi-Levi, David and Zhao, Jinglong},
  journal={Management Science},
  volume={69},
  number={7},
  pages={3759--3777},
  year={2023},
  publisher={INFORMS}
}

@article{xiong2023optimal,
  title={Optimal experimental design for staggered rollouts},
  author={Xiong, Ruoxuan and Athey, Susan and Bayati, Mohsen and Imbens, Guido},
  journal={Management Science},
  year={2023},
  publisher={INFORMS}
}

@article{hahn2011adaptive,
  title={Adaptive experimental design using the propensity score},
  author={Hahn, Jinyong and Hirano, Keisuke and Karlan, Dean},
  journal={Journal of Business \& Economic Statistics},
  volume={29},
  number={1},
  pages={96--108},
  year={2011},
  publisher={Taylor \& Francis}
}

@inproceedings{atan2019sequential,
  title={Sequential patient recruitment and allocation for adaptive clinical trials},
  author={Atan, Onur and Zame, William R and Schaar, Mihaela},
  booktitle={The 22nd International Conference on Artificial Intelligence and Statistics},
  pages={1891--1900},
  year={2019},
  organization={PMLR}
}

@article{greenhill2020bayesian,
  title={Bayesian optimization for adaptive experimental design: A review},
  author={Greenhill, Stewart and Rana, Santu and Gupta, Sunil and Vellanki, Pratibha and Venkatesh, Svetha},
  journal={IEEE access},
  volume={8},
  pages={13937--13948},
  year={2020},
  publisher={IEEE}
}

@article{kato2020efficient,
  title={Efficient adaptive experimental design for average treatment effect estimation},
  author={Kato, Masahiro and Ishihara, Takuya and Honda, Junya and Narita, Yusuke},
  journal={arXiv preprint arXiv:2002.05308},
  year={2020}
}

@article{yang2017framework,
  title={A framework for multi-a (rmed)/b (andit) testing with online fdr control},
  author={Yang, Fanny and Ramdas, Aaditya and Jamieson, Kevin G and Wainwright, Martin J},
  journal={Advances in Neural Information Processing Systems},
  volume={30},
  year={2017}
}

@inproceedings{yao2021power,
  title={Power constrained bandits},
  author={Yao, Jiayu and Brunskill, Emma and Pan, Weiwei and Murphy, Susan and Doshi-Velez, Finale},
  booktitle={Machine Learning for Healthcare Conference},
  pages={209--259},
  year={2021},
  organization={PMLR}
}

@inproceedings{erraqabi2017trading,
  title={Trading off rewards and errors in multi-armed bandits},
  author={Erraqabi, Akram and Lazaric, Alessandro and Valko, Michal and Brunskill, Emma and Liu, Yun-En},
  booktitle={Artificial Intelligence and Statistics},
  pages={709--717},
  year={2017},
  organization={PMLR}
}

@article{guha2013nearly,
  title={(Nearly) optimal algorithms for private online learning in full-information and bandit settings},
  author={Guha Thakurta, Abhradeep and Smith, Adam},
  journal={Advances in Neural Information Processing Systems},
  volume={26},
  year={2013}
}

@inproceedings{charisopoulos2023robust,
  title={Robust and private stochastic linear bandits},
  author={Charisopoulos, Vasileios and Esfandiari, Hossein and Mirrokni, Vahab},
  booktitle={International Conference on Machine Learning},
  pages={4096--4115},
  year={2023},
  organization={PMLR}
}

@article{agarwal2021causal,
  title={Causal inference with corrupted data: Measurement error, missing values, discretization, and differential privacy},
  author={Agarwal, Anish and Singh, Rahul},
  journal={arXiv preprint arXiv:2107.02780},
  year={2021}
}

@inproceedings{kusner2016private,
  title={Private causal inference},
  author={Kusner, Matt J and Sun, Yu and Sridharan, Karthik and Weinberger, Kilian Q},
  booktitle={Artificial Intelligence and Statistics},
  pages={1308--1317},
  year={2016},
  organization={PMLR}
}

@article{lee2013neighborhood,
  title={Neighborhood social capital and social learning for experience attributes of products},
  author={Lee, Jae Young and Bell, David R},
  journal={Marketing Science},
  volume={32},
  number={6},
  pages={960--976},
  year={2013},
  publisher={INFORMS}
}

@article{simple-regret-instance-lowerbound,
author = {Jean-Yves Audibert and Alexandre B. Tsybakov},
title = {{Fast learning rates for plug-in classifiers}},
volume = {35},
journal = {The Annals of Statistics},
number = {2},
publisher = {Institute of Mathematical Statistics},
pages = {608 -- 633},
year = {2007},
doi = {10.1214/009053606000001217},
URL = {https://doi.org/10.1214/009053606000001217}
}

\section{Appendix: Proof of Inequality \ref{eq: minimal regret v.s. margin condition}} \label{proof: minimal regret v.s. margin condition}

For any $\nu \in V(\alpha,1)$, we have
\begin{align*}
\E_X\!\left[\min\{n|\Delta(X)|,|\Delta(X)|^{-(1+d)}\}\right]
&\leqslant
n^{\frac{1+d}{2+d}}\P\!\left(|\Delta(X)|\leqslant n^{-\frac{1}{2+d}}\right)
+\E\!\left[|\Delta(X)|^{-(1+d)}\mathbf{1}\{|\Delta(X)|\geqslant n^{-\frac{1}{2+d}}\}\right] \\
&\leqslant
n^{\frac{1+d}{2+d}}\cdot n^{-\frac{\alpha}{2+d}}
+\int_{n^{-\frac{1}{2+d}}}^{1}\delta^{-(1+d)}p(|\Delta(X)|=\delta)\,d\delta \\
&\leqslant
n^{\frac{(1-\alpha)+d}{2+d}}
+\int_{n^{-\frac{1}{2+d}}}^{1}\delta^{-(1+d)}\cdot D\alpha\delta^{\alpha-1}\,d\delta \\
&=
n^{\frac{(1-\alpha)+d}{2+d}}
+O\!\left(n^{\frac{1}{2+d}((1+d)-\alpha)}\right) \\
&=
O\!\left(n^{\frac{(1-\alpha)+d}{2+d}}\right).
\end{align*}
Moreover, this rate is tight, since
\[
\sup_{\nu\in V(\alpha,1)}
\E_X\!\left[\min\{n|\Delta(X)|,|\Delta(X)|^{-(1+d)}\}\right]
=
O\!\left(n^{\frac{(1-\alpha)+d}{2+d}}\right).
\]

\section{Appendix: Proof of Theorem \ref{thm-lower}}
The lower bound
\[
R_{n,\nu'}(\pi) \geqslant R_{min}
\]
is by the minimal regret assumption \ref{ass: minimal regret assumption}. We therefore focus on establishing
\[
R_{n,\nu'}(\pi) \geqslant 
\softO\left(
E_{n,\nu'}^{-\frac{2+d}{2}}(\pi)\cdot 
\E_X\!\left[|\Delta_{\nu'}(X)|^{\frac{2}{4+d}}\right]^{\frac{4+d}{2}}
\right).
\]

\paragraph{Adaptive partition.}
Fix an instance $\nu$. We construct an adaptive dyadic partition of $\X=[0,1]^d$ according to the local difficulty of $\nu$.

\begin{definition}[Adaptive partition $\calP_\nu$] \label{def: Adaptive Partition}
The partition $\calP_\nu$ is generated recursively as follows.
\begin{enumerate}
\item \textbf{Initialization.} Initialize with the root cube $S_0=[0,1]^d$ at level $k=0$.
\item \textbf{Splitting rule.} For a cube $S$ at level $k$ with side length $a_k=2^{-k}$ and center $x_c$, evaluate $|\Delta_\nu(x_c)|$.
\begin{itemize}
    \item If $|\Delta_\nu(x_c)| \geqslant 3L\Big(\tfrac{\sqrt d\,a_k}{2}\Big)$, declare $S$ \textsc{Terminal} and add it to $\calP_\nu$.
    \item Otherwise, subdivide $S$ into $2^d$ dyadic subcubes of level $k+1$ and recurse on each subcube.
\end{itemize}
\item \textbf{Depth truncation.} The recursion stops once a cube is terminal or when the level reaches
\[
\kmax=\Big\lceil \tfrac{1}{2+d}\log_2 n \Big\rceil,
\]
in which case all remaining (non-terminal) cubes at level $\kmax$ are declared terminal and added to $\calP_\nu$.
\end{enumerate}
\end{definition}

\begin{figure}[t]
    \centering
    \includegraphics[width=\textwidth]{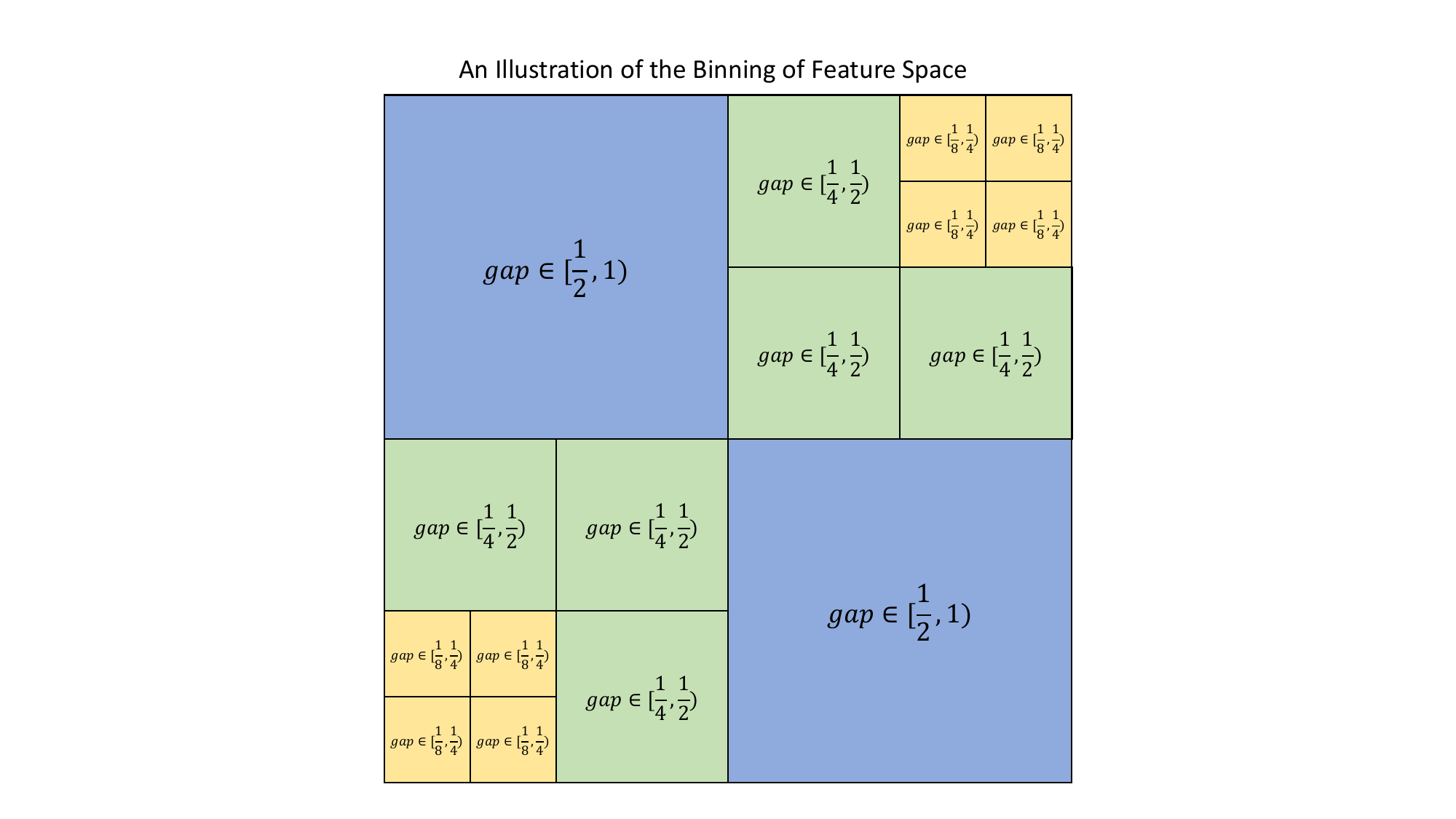}
    \caption{An Illustration of the Binning of Feature Space}
    \label{fig:Adaptive_Partition_Illustration}
\end{figure}

\medskip
\paragraph{Basic geometric and Lipschitz consequences.}
Let $C_k$ denote the number of (terminal) cubes in $\calP_\nu$ at level $k$ and define $p_k:=C_k a_k^d$. Set $\Delta_k:=a_k$.
For any level-$k$ cube $S$ with center $x_c$, we have the geometric bound
\[
\|x-x_c\|_2 \leqslant \frac{\sqrt d\,a_k}{2}\qquad \forall\,x\in S,
\]
since each coordinate differs from the center by at most $a_k/2$.
Because each arm mean is $L$-Lipschitz, the gap $\Delta_\nu=f_\nu^{(2)}-f_\nu^{(1)}$ is $2L$-Lipschitz, hence for all $x\in S$,
\[
|\Delta_\nu(x)-\Delta_\nu(x_c)|
\leqslant 2L\|x-x_c\|_2
\leqslant 2L\Big(\tfrac{\sqrt d\,a_k}{2}\Big).
\]
Therefore, if $S$ is declared terminal at level $k<\kmax$, then for every $x\in S$,
\[
|\Delta_\nu(x)|
\geqslant |\Delta_\nu(x_c)|-|\Delta_\nu(x)-\Delta_\nu(x_c)|
\geqslant 3L\Big(\tfrac{\sqrt d\,a_k}{2}\Big)-2L\Big(\tfrac{\sqrt d\,a_k}{2}\Big)
= \frac{1}{2}L\sqrt d\,a_k>0,
\]
so $\Delta_\nu(\cdot)$ has constant sign on $S$, and hence the optimal arm is constant on $S$.

Moreover, on any $X$ in a level-$k$ cube $S$, the parent cube $S'$ of $S$ in level $k-1$ \emph{fails} the terminal test (and thus is split), we have
$|\Delta_\nu(x'_c)|<3L(\tfrac{\sqrt d\,a_{k-1}}{2})$, and consequently for all $x\in S$,
\[
|\Delta_\nu(x)|
\leqslant |\Delta_\nu(x'_c)|+|\Delta_\nu(x)-\Delta_\nu(x'_c)|
< 3L\Big(\tfrac{\sqrt d\,a_{k-1}}{2}\Big)+2L\Big(\tfrac{\sqrt d\,a_{k-1}}{2}\Big)
= 5L\sqrt d\,a_k
\].

For $X\in\X$, let $k(X)$ denote the level of the (unique) terminal cube in $\calP_\nu$ that contains $X$. And therefore as a conclusion of above, we have:
\begin{align}\label{eq: gap control}
    \forall X\in \mathcal{X}, |\Delta(X)|\in [\frac{1}{2}L\sqrt d\,a_k,\, 5L\sqrt d\,a_k]\quad &\text{ for }\quad 0<k(X)<\kmax;\\
    \forall X\in \mathcal{X}, |\Delta(X)| < 5L\sqrt d\,a_k \quad &\text{ for }\quad k(X)=\kmax;
\end{align}

\paragraph{Hardest scale and regret lower bound.}
Define
\begin{align}\label{def: k_star}
    k^*\in \text{argmax}_{k\leqslant\kmax} p_k a_k^{\frac{2}{4+d}}.
\end{align}

By the minimal regret assumption,
\[
R \coloneqq R_{n,\nu}(\pi)\geqslant R_{\min}
\geqslant p_{k^*}\Delta_{k^*}^{-(1+d)}
=\frac{C_{k^*}}{\Delta_{k^*}}.
\]

\paragraph{Construction of perturbed instances.}
We now construct a family of locally perturbed instances around a baseline instance, following a standard packing argument, while preserving the smoothness and margin structure induced by the adaptive partition.

\medskip
\textbf{Baseline instance.}
Define an auxiliary instance $\nu'$ by setting the first-arm mean identically equal to one,
\[
f^{(1)}_{\nu'}(x)\equiv 1 \qquad \forall\,x\in\X.
\]
Fix a level $k$ and consider any level-$k$ cube $S$ from the adaptive partition, with center $x_c$ and side length $a_k=\Delta_k$.
On the central region of $S$, define the second-arm mean by
\[
f^{(2)}_{\nu'}(x)
= 1-\tfrac{1}{2}L\Delta_k
\qquad \text{for all } x \text{ such that } \|x-x_c\|_\infty\leqslant \tfrac{a_k}{2}.
\]
Outside this region, extend $f^{(2)}_{\nu'}$ smoothly so that $f^{(2)}_{\nu'}$ remains bounded in $[0,1]$ and is $L$-Lipschitz on $\X$.
Such an extension exists by standard smooth interpolation arguments and ensures that the resulting instance $\nu'$ is well defined.

\medskip
\textbf{Baseline regret scale.}
Let $k^*$ denote the critical level defined in \ref{def: k_star}, and recall that $C_{k^*}$ denotes the number of level-$k^*$ cubes in the adaptive partition.
Define the benchmark regret level
\[
R'_{\min} \;:=\; \frac{C_{k^*}}{\Delta_{k^*}}.
\]
By construction, on each level-$k^*$ cube the gap between the two arms is of order $\Delta_{k^*}$ on a set of non-negligible volume. Consequently, any policy $\pi$ must incur expected cumulative regret at least $R'_{\min}$ on $\nu'$, that is,
\[
R_{n,\nu'}(\pi)\geqslant R_{\min}(\nu')
\geqslant p_{k^*}\Delta_{k^*}^{-(1+d)}
\asymp\frac{C_{k^*}}{\Delta_{k^*}}= R'_{\min},
\]
uniformly over all admissible policies $\pi$.

\medskip
\textbf{Subdivision at the critical level.}
For each level-$k^*$ cube, we further subdivide its central region into $M^d$ equal dyadic subcubes, where
\[
M
:= \Big\lfloor \Big(\tfrac{R}{R'_{\min}}\Big)^{\frac{1}{2+d}} \Big\rfloor
\;\geqslant\;1.
\]
Denote by $\{\mathrm{Sub}_j\}_{j=1}^{C_{k^*}M^d}$ the resulting collection of subcubes across all level-$k^*$ cubes. Each subcube has side length $a_{k^*}/M$ and is contained in the interior of its parent cube.

\medskip
\textbf{Perturbation amplitude.}
Define the perturbation magnitude
\[
\delta
:= \frac{1}{(\log n)^2}
\sqrt{\frac{C_{k^*}\Delta_{k^*}M^d}{R}}
\;\asymp\;
\frac{\Delta_{k^*}}{(\log n)^2}
\Big(\frac{R'_{\min}}{R}\Big)^{\frac{1}{2+d}}
\;\leqslant\;
\frac{\Delta_{k^*}}{(\log n)^2}.
\]
The upper bound ensures that the perturbations are asymptotically small relative to the baseline gap at level $k^*$, and the logarithmic factor provides additional slack required later in the information-theoretic argument.

\medskip
\textbf{Local bump functions.}
Let $\psi:\R^d\to\R$ be a fixed smooth function supported on $[-\tfrac12,\tfrac12]^d$ with $\|\psi\|_\infty\leqslant 1$.
For each subcube $\mathrm{Sub}_j$ with center $x^{\mathrm{sub}}_j$, define the localized bump
\[
\phi_j(x)
:= \delta\,
\psi\!\left(\frac{x-x^{\mathrm{sub}}_j}{a_{k^*}/M}\right).
\]
By construction, $\phi_j$ is supported on $\mathrm{Sub}_j$ and has amplitude at most $\delta$, while its smoothness constants scale proportionally with $L$.

\medskip
\textbf{Perturbed instances.}
For each sign vector $\varepsilon=(\epsilon_j)_{j=1}^{C_{k^*}M^d}\in\{\pm1\}^{C_{k^*}M^d}$, define an instance $\nu_\varepsilon$ by
\begin{align*}
f^{(1)}_{\nu_\varepsilon}(x) &= f^{(1)}_{\nu'}(x), \\
f^{(2)}_{\nu_\varepsilon}(x)
&= f^{(2)}_{\nu'}(x)
+ \sum_{j=1}^{C_{k^*}M^d} \epsilon_j\,\phi_j(x).
\end{align*}
That is, $f^{(2)}_{\nu_\varepsilon}$ coincides with the baseline function $f^{(2)}_{\nu'}$ outside the union of subcubes, while within each subcube $\mathrm{Sub}_j$ it is locally perturbed upward or downward according to the sign $\epsilon_j$.
By the choice of $\delta$ and the smoothness of $\psi$, all instances $\nu_\varepsilon$ remain $L$-Lipschitz and hence belong to the same model class.

\paragraph{Local comparability of gaps.}
We record a simple but important regularity property of the adaptive partition, which ensures that gap magnitudes remain comparable across neighboring cubes and across the perturbed instances.

Fix a point $x\in\X$, and let $S_k$ denote the level-$k$ terminal cube of the adaptive partition $\calP_\nu$ that contains $x$.
Let $S_\ell$ be any neighboring terminal cube (sharing a common face, edge, or vertex with $S_k$), and let $\ell$ denote its level. Then considering their shared face, (\ref{eq: gap control}) implies that:
\[
|k-\ell|\;\leqslant\;\Big\lceil \log_2 5 \Big\rceil.
\]

Combining the bounded level difference $|k-\ell|$ with the fact that $\Delta_k=2^{-k}$ varies geometrically in $k$, we conclude that for all $x\in\X$,
\[
\frac{1}{10}(\sqrt d)
\;\leqslant\;
\frac{|\Delta_\nu(x)|}{|\Delta_{\nu_\varepsilon}(x)|}
\;\leqslant\;
100(\sqrt d).
\]
These bounds ensure that the gap functions under $\nu$ and $\nu_\varepsilon$ are locally comparable up to fixed dimension-dependent constants, a fact that will be used repeatedly in the subsequent regret and information-theoretic arguments.

\paragraph{Information-theoretic argument.}
We now relate the regret constraint to a lower bound on the estimation error via a standard change-of-measure argument.

For each subcube $\mathrm{Sub}_j$, define the (random) number of pulls of arm~2 within this region by
\[
T_j \;:=\; \sum_{t=1}^n \mathbf{1}\{X_t\in \mathrm{Sub}_j,\ a_t=2\}.
\]
By construction of the baseline instance at level $k^*$, pulling arm~2 on $\mathrm{Sub}_j$ incurs an instantaneous regret of order $\Delta_{k^*}$. Therefore, the expected cumulative regret contributed by all subcubes satisfies
\[
\sum_{j} \Delta_{k^*}\,\E[T_j] \;\leqslant\; R,
\]
which implies
\[
\sum_j \E[T_j] \;\leqslant\; \frac{R}{\Delta_{k^*}}.
\]
Since there are $C_{k^*}M^d$ subcubes in total, a simple averaging argument yields that at least half of the indices $j$ must satisfy
\[
\E[T_j] \;\leqslant\; \frac{2R}{C_{k^*}M^d\Delta_{k^*}}.
\]

Fix such an index $j$. Let $\varepsilon^j$ denote the sign vector obtained from $\varepsilon$ by flipping only the $j$-th coordinate.
The instances $\nu_\varepsilon$ and $\nu_{\varepsilon^j}$ differ only on $\mathrm{Sub}_j$, and only through the sign of the local perturbation $\phi_j$.
Using the standard expression for the Kullback--Leibler divergence between Gaussian (or unit-variance sub-Gaussian) observation models with identical noise distributions, we obtain
\begin{align*}
D_{\mathrm{KL}}\!\left(\mathbb{P}^{(n)}_{\nu_\varepsilon}\,\middle\|\,\mathbb{P}^{(n)}_{\nu_{\varepsilon^j}}\right)
&=
\frac12 \E_{\nu_\varepsilon}\!\left[
\sum_{t=1}^n
\big(f^{(a_t)}_{\nu_\varepsilon}(X_t)-f^{(a_t)}_{\nu_{\varepsilon^j}}(X_t)\big)^2
\right].
\end{align*}
The summand is nonzero only when $X_t\in \mathrm{Sub}_j$ and $a_t=2$, in which case the difference between the two instances equals $2\phi_j(X_t)$.
Using $\|\phi_j\|_\infty\leqslant\delta$, we therefore have
\begin{align*}
D_{\mathrm{KL}}\!\left(\mathbb{P}^{(n)}_{\nu_\varepsilon}\,\middle\|\,\mathbb{P}^{(n)}_{\nu_{\varepsilon^j}}\right)
&\leqslant
\frac12 \E_{\nu_\varepsilon}\!\left[
\sum_{t=1}^n
\mathbf{1}\{X_t\in \mathrm{Sub}_j,\ a_t=2\}\,(2\phi_j(X_t))^2
\right] \\
&\leqslant
2\,\E\!\left[
\sum_{t=1}^n
\mathbf{1}\{X_t\in \mathrm{Sub}_j,\ a_t=2\}\,\phi_j(X_t)^2
\right] \\
&=
2\delta^2\,\E[T_j].
\end{align*}
For the indices $j$ under consideration, this yields
\[
D_{\mathrm{KL}}\!\left(\mathbb{P}^{(n)}_{\nu_\varepsilon}\,\middle\|\,\mathbb{P}^{(n)}_{\nu_{\varepsilon^j}}\right)
\;\leqslant\;
2\delta^2\cdot \frac{2R}{C_{k^*}M^d\Delta_{k^*}}
\;\leqslant\;
\frac{4}{\log n}
\;\leqslant\; 4,
\]
where the last inequality follows from the definition of $\delta$ and holds for $n$ sufficiently large.

Applying the Bretagnolle--Huber inequality, we obtain the total variation bound
\[
1-d_{\mathrm{TV}}\!\left(\mathbb{P}^{(n)}_{\nu_\varepsilon},\mathbb{P}^{(n)}_{\nu_{\varepsilon^j}}\right)
\;\geqslant\;
1-\sqrt{1-e^{-D_{\mathrm{KL}}}}
\;\geqslant\;
1-\sqrt{1-e^{-4}}.
\]

This indistinguishability implies a lower bound on the estimation error on $\mathrm{Sub}_j$.
Let $E^j_{n,\nu}(\pi)$ denote the squared estimation error of policy $\pi$ on $\mathrm{Sub}_j$ under instance $\nu$.
By standard testing arguments,
\[
E^j_{n,\nu_\varepsilon}(\pi)+E^j_{n,\nu_{\varepsilon^j}}(\pi)
\;\geqslant\;
\big(1-\sqrt{1-e^{-4}}\big)\,\delta^2
\;=\; \Theta(\delta^2),
\]
and hence, by symmetry over $\varepsilon$,
\[
\E_\varepsilon\!\left[E^j_{n,\nu_\varepsilon}(\pi)\right]
\;\geqslant\; \Theta(\delta^2).
\]

Since the above argument applies to at least half of the $C_{k^*}M^d$ subcubes, summing over these indices yields
\[
\E_\varepsilon\!\left[E_{n,\nu_\varepsilon}(\pi)\right]
\;\geqslant\;
\Theta\!\big(p_{k^*}\,\delta^2\big)
\;=\;
\softO\!\left(
R^{-\frac{2}{2+d}}\,
C_{k^*}^{\frac{4+d}{2+d}}\,
a_{k^*}^{1+d+\frac{d}{2+d}}
\right),
\]
where we used $p_{k^*}=C_{k^*}a_{k^*}^d$ and the definition of $\delta$.
In particular, there exists at least one sign vector $\varepsilon$ such that
\[
E_{n,\nu_\varepsilon}(\pi)
\;\geqslant\;
\softO\!\left(
R^{-\frac{2}{2+d}}\,
C_{k^*}^{\frac{4+d}{2+d}}\,
a_{k^*}^{1+d+\frac{d}{2+d}}
\right),
\]
which completes the information-theoretic lower bound.

\paragraph{Completion of the proof.}
We conclude by combining the preceding information-theoretic bound with a case analysis on the realized regret of the policy.

\medskip
\textbf{Case 1: Large regret.}
Suppose first that
\[
R_{n,\nu_\varepsilon}(\pi)\;\geqslant\;\tfrac12 R.
\]
Multiplying both sides of the lower bound on the estimation error derived above by
$R_{n,\nu_\varepsilon}(\pi)^{\frac{2}{2+d}}$, and using the assumption
$R_{n,\nu_\varepsilon}(\pi)\asymp R$ up to constants, we obtain
\begin{align*}
R_{n,\nu_\varepsilon}(\pi)^{\frac{2}{2+d}}\,E_{n,\nu_\varepsilon}(\pi)
&\;\geqslant\;
\softO\!\left(
C_{k^*}^{\frac{4+d}{2+d}}\,
a_{k^*}^{1+d+\frac{d}{2+d}}
\right).
\end{align*}
Recalling that $p_{k^*}=C_{k^*}a_{k^*}^d$ and rearranging powers of $a_{k^*}$, this bound can be rewritten as
\begin{align*}
C_{k^*}^{\frac{4+d}{2+d}}\,a_{k^*}^{1+d+\frac{d}{2+d}}
&=
\left(p_{k^*}\,a_{k^*}^{\frac{2}{4+d}}\right)^{\frac{4+d}{2+d}},
\end{align*}
up to logarithmic factors.
By the definition (\ref{def: k_star}) of $k^*$,
\[
\E_X\!\left[|\Delta_\nu(X)|^{\frac{2}{4+d}}\right]\leqslant \kmax\cdot p_{k^*}\,a_{k^*}^{\frac{2}{4+d}}=\softO\left(p_{k^*}\,a_{k^*}^{\frac{2}{4+d}}\right).
\]
Combining the above displays yields
\[
R_{n,\nu_\varepsilon}(\pi)^{\frac{2}{2+d}}\,E_{n,\nu_\varepsilon}(\pi)
\;\geqslant\;
\softO\!\left(
\E_X\!\left[|\Delta_\nu(X)|^{\frac{2}{4+d}}\right]^{\frac{4+d}{2+d}}
\right).
\]

\medskip
\textbf{Case 2: Small regret.}
If instead
\[
R_{n,\nu_\varepsilon}(\pi)\;<\;\tfrac12 R,
\]
we replace the current instance $\nu$ by the perturbed instance $\nu_\varepsilon$ and repeat the above construction.
By design, each iteration reduces the regret level by a constant factor while maintaining the lower bound
$R_{n,\nu_\varepsilon}(\pi)\geqslant R'_{\min}$.
Hence, this iterative procedure can be repeated at most $O(\log n)$ times before termination.

At each iteration, the additional perturbation to the gap function is bounded in magnitude by
$\Delta_{k^*}/(\log n)^2$.
Therefore, after at most $\log n$ iterations, the cumulative perturbation is at most
$\Delta_{k^*}/\log n$, which is negligible compared to the baseline gap at level $k^*$.
In particular, the adaptive partition $\calP_\nu$ and all associated quantities
($k^*$, $C_{k^*}$, $a_{k^*}$, and $p_{k^*}$) remain unchanged throughout the procedure.

\medskip
\textbf{Conclusion.}
Consequently, after at most $\log n$ iterations, there exists an instance $\nu'$ in the model class such that
\[
R_{n,\nu'}(\pi)^{\frac{2}{2+d}}\,E_{n,\nu'}(\pi)
\;\geqslant\;
\softO\!\left(
\E_X\!\left[|\Delta_\nu(X)|^{\frac{2}{4+d}}\right]^{\frac{4+d}{2+d}}
\right).
\]
Equivalently, rearranging terms yields the desired trade-off bound
\[
R_{n,\nu'}(\pi)
\;\geqslant\;
\softO\!\left(
E_{n,\nu'}(\pi)^{-\frac{2+d}{2}}\cdot
\E_X\!\left[|\Delta_{\nu'}(X)|^{\frac{2}{4+d}}\right]^{\frac{4+d}{2}}
\right),
\]
which completes the proof of the second conclusion of the theorem.

\section{Appendix: Proof of Theorem \ref{thm_upper_noprivacy}}
Suppose the noise variables are $\sigma$-sub-Gaussian for all $x\in\X$ and $a\in\{0,1\}$.

\paragraph{A high-probability event.}
We begin by defining a “good event” under which all empirical gap estimators are uniformly accurate across all stages and all cubes generated by the algorithm.
Specifically, define
\begin{align*}
\mathcal{E}
:=
\Bigl\{
\forall\,1\leqslant i\leqslant H,\ \forall\,S,\ 
\bigl|\Delta(S)-\hat{\Delta}_i(S)\bigr|
\leqslant
2\sigma\sqrt{\log n}\,\Delta_i
\Bigr\},
\end{align*}
where $\Delta(S)$ denotes the true average gap on cube $S$, $\hat{\Delta}_i(S)$ is its empirical estimate computed at step $i$, and $\Delta_i$ is the scale parameter associated with step $i$.

We first control the deviation probability for a fixed pair $(i,S)$.
Conditioning on the covariates and arm assignments, $\hat{\Delta}_i(S)$ is an average of independent $\sigma$-sub-Gaussian noise terms.
Therefore, by the standard Hoeffding (sub-Gaussian) concentration inequality, for any fixed cube $S$ and step $i$,
\begin{align*}
\mathbb{P}\!\left(
\bigl|\Delta(S)-\hat{\Delta}_i(S)\bigr|
>
2\sigma\sqrt{\log n}\,\Delta_i
\right)
&\leqslant
2\exp\!\left(
-\frac{(2\sigma\sqrt{\log n}\,\Delta_i)^2}{2\sigma^2\Delta_i^2}
\right) \\
&= 2e^{-2\log n}\\
&= \frac{2}{n^2}.
\end{align*}

We now take a union bound over all estimators involved in the algorithm.
At step $i$, the partition consists of at most $2^{di}$ cubes.
Since the total number of steps satisfies
\[
H \leqslant \frac{2}{2+d}\log_2 n,
\]
the total number of pairs $(i,S)$ over which the event $\mathcal{E}$ is defined is at most
\[
\sum_{i=1}^H 2^{di}
\;\leqslant\;
H\,2^{dH}
\;\leqslant\;
\frac{2}{2+d}\log_2 n \cdot n^{\frac{d}{2+d}}.
\]
Applying a union bound yields
\begin{align*}
\mathbb{P}(\mathcal{E}^c)
&\leqslant
\frac{2}{n^2}
\cdot
\frac{2}{2+d}\log_2 n \cdot n^{\frac{d}{2+d}}.
\end{align*}
For sufficiently large $n$, the right-hand side is bounded by $1/n$, and hence
\[
\mathbb{P}(\mathcal{E})
\;\geqslant\;
1-\frac{1}{n}.
\]

In the remainder of the proof, we condition on the event $\mathcal{E}$, which holds with high probability, and analyze the behavior of the algorithm deterministically under this event.

\paragraph{First-phase Regret.}
We now upper bound the cumulative regret of the first phase, starting with its contribution on the high-probability event $\mathcal{E}$ defined above.

\medskip
\textbf{Correctness of arm elimination.}
On the event $\mathcal{E}$, all empirical gap estimates are uniformly accurate at their corresponding scales.
In particular, whenever a cube is declared \emph{terminal} and an arm is labeled as “optimal,” the sign of the estimated gap coincides with that of the true gap.
Hence, under $\mathcal{E}$, the algorithm never commits to a suboptimal arm.
As a consequence, during Phase~I, regret is incurred only by those cubes that have not yet been declared terminal and in which the algorithm continues to explore both arms.

\medskip
\textbf{Gap control on non-terminal cubes.}
Fix a step $k$ and consider a cube $S$ at level $k$ that has not yet terminated.
By the stopping rule of the algorithm, such a cube must satisfy
\[
|\hat{\Delta}_k(S)| \;\leqslant\; \log n \cdot \Delta_k.
\]
On the event $\mathcal{E}$ and for $n$ sufficiently large, the deviation bound implies
\begin{align*}
|\Delta(S)|
&\leqslant
|\hat{\Delta}_k(S)| + \bigl|\Delta(S)-\hat{\Delta}_k(S)\bigr| \\
&\leqslant
\log n \cdot \Delta_k + 2\sigma\sqrt{\log n}\,\Delta_k\\
&\leqslant
\frac{3}{2}\log n \cdot \Delta_k,
\end{align*}
where the last inequality follows by absorbing constants into $\log n$.

Next, recall that $S$ has side length $a_k=2^{-k}$.
By Lipschitz continuity of the gap function and the geometry of the cube, for any $x\in S$,
\begin{align*}
|\Delta(x)|
&\leqslant
|\Delta(S)| + L\|x-x_S\| \\
&\leqslant
|\Delta(S)| + L\sqrt{d}\,a_k \\
&\leqslant
2\log n \cdot \Delta_k,
\end{align*}
where $x_S$ denotes the center of $S$, and we again absorb fixed constants into $\log n$.

\medskip
\textbf{Termination depth at a point.}
The preceding bound implies that a point $x\in\X$ cannot remain in a non-terminal cube once the scale $\Delta_k$ becomes sufficiently small relative to $|\Delta(x)|$.
More precisely, define
\begin{align*}
k_x
&:=
\min\bigl\{k:\ |\Delta(x)|>2\log n\cdot \Delta_k\bigr\}.
\end{align*}
Since $\Delta_k=2^{-k}$, a direct calculation yields
\begin{align*}
k_x
&=
\Bigl\lfloor
\log_2\log n + 1 - \log_2|\Delta(x)|
\Bigr\rfloor + 1.
\end{align*}
Thus, after at most $k_x$ refinement steps, the cube containing $x$ must be declared terminal.

\medskip
\textbf{Regret accumulated before termination.}
Recall that the batch lengths grow geometrically with ratio $r=2^{d+2}$.
Hence, the total number of exploration rounds incurred at a point $x$ before termination is at most
\[
r + r^2 + \cdots + r^{k_x}
\;\leqslant\;
r^2 \min\{r^{k_x-1},\,n\}.
\]
Using the expression for $k_x$ and the definition of $r$, we obtain
\begin{align*}
r^{k_x-1}
&\leqslant
2^{(2+d)\left(\log_2\log n + 1 - \log_2|\Delta(x)|\right)} \\
&\leqslant
(2\log n)^{2+d}\,|\Delta(x)|^{-(2+d)}.
\end{align*}

\medskip
\textbf{Integration over the covariate space.}
Accounting for the event $\mathcal{E}$ and its complement, the regret incurred during the first phase satisfies
\begin{align*}
R_1
&\leqslant
\int_\X |\Delta(x)|\,
\frac{1}{2}\min\{r+r^2+\cdots+r^{k_x},\,n/2\}
\,d\mathcal{P}(x)
+ \mathbb{P}(\mathcal{E}^c)\cdot \frac{n}{2}\E_X[|\Delta(X)|] \\
&\leqslant
r^2 \int_\X |\Delta(x)|\min\{r^{k_x-1},\,n\}\,d\mathcal{P}(x)
+ \frac{1}{2}\E_X[|\Delta(X)|].
\end{align*}
Substituting the bound on $r^{k_x-1}$ gives
\begin{align*}
R_1
&\leqslant
r^2(2\log n)^{2+d}
\int_\X
\min\{|\Delta(x)|^{-(1+d)},\,n|\Delta(x)|\}
\,d\mathcal{P}(x)
+ \frac{1}{2}\E_X[|\Delta(X)|].
\end{align*}
Absorbing polynomial and logarithmic factors into the $\softO(\cdot)$ notation, we conclude that
\begin{align*}
R_1
&\leqslant
\softO\!\left(
\E_X\!\left[\min\{|\Delta(X)|^{-(1+d)},\,n|\Delta(X)|\}\right]
\right)
=
R_{\min}.
\end{align*}

\paragraph{Second-phase regret.}
We now bound the regret incurred during Phase~II, again conditioning on the high-probability event $\mathcal{E}$.

\medskip
\textbf{Gap magnitude on terminal cubes.}
Consider a cube $S$ that becomes terminal at step $k$.
By the termination rule of Phase~I, this means that
\[
|\hat{\Delta}_k(S)|>\log n\cdot \Delta_k
\qquad
\text{(or the symmetric condition with negative sign).}
\]
Fix any $x\in S$.
On the event $\mathcal{E}$, the estimation error bound and Lipschitz continuity imply
\begin{align*}
|\Delta(x)|
&\geqslant
|\Delta(S)| - L\|x-x_S\| \\
&\geqslant
|\hat{\Delta}_k(S)|
- \bigl|\hat{\Delta}_k(S)-\Delta(S)\bigr|
- L\sqrt{d}\,a_k \\
&\geqslant
\log n\cdot \Delta_k
- 2\sigma\sqrt{\log n}\,\Delta_k
- L\sqrt{d}\,\Delta_k \\
&>
\frac{1}{2}\log n\cdot \Delta_k,
\end{align*}
where $x_S$ denotes the center of $S$, and the last inequality holds for all sufficiently large $n$ after absorbing constants into $\log n$.

We also require an upper bound on the gap magnitude within $S$.
Let $S'$ denote the parent cube of $S$ at level $k-1$.
Since $S'$ did not terminate at step $k-1$, its empirical gap satisfies
$|\hat{\Delta}_{k-1}(S')|\leqslant \log n\cdot \Delta_{k-1}$.
Applying the event $\mathcal{E}$ and Lipschitz continuity yields, for any $x\in S$,
\begin{align*}
|\Delta(x)|
&\leqslant
|\Delta(S')| + L\sqrt{d}\,a_{k-1} \\
&\leqslant
|\hat{\Delta}_{k-1}(S')|
+ 2\sigma\sqrt{\log n}\,\Delta_{k-1}
+ L\sqrt{d}\,\Delta_{k-1} \\
&<
3\log n\cdot \Delta_k,
\end{align*}
where we used $\Delta_{k-1}=2\Delta_k$ and again absorbed constants into the logarithmic factor.
The same upper bound applies to cubes that remain non-terminal until the final step $H$.

\medskip
\textbf{Regret decomposition in Phase~II.}
During Phase~II, the algorithm predominantly exploits the arm labeled as optimal within each terminal cube, while deliberately randomizing with probability $q_k$ on level-$k$ cubes ($k<H$) to facilitate estimation.
For a point $x$ belonging to a level-$k$ terminal cube, the instantaneous regret incurred by this randomization is of order $|\Delta(x)|$, which is bounded by $3\log n\cdot \Delta_k$ as shown above.
Therefore, summing over all levels and integrating over the covariate distribution, the expected Phase~II regret satisfies
\begin{align*}
R_2
&\leqslant
n\!\left(
\sum_{k=1}^{H-1} p_k\, q_k \cdot 3\log n\cdot \Delta_k
+
\E_X\!\left[
|\Delta(X)|\,\mathbf{1}_{\{|\Delta(X)|<3\log n\cdot \Delta_H\}}
\right]
\right)
+ \E_X[|\Delta(X)|].
\end{align*}
Here, the second term accounts for points lying in cubes that do not terminate before the final level $H$, and the final additive term captures the negligible contribution from the complement of $\mathcal{E}$.

\medskip
\textbf{Final bound.}
Recalling the definition
\[
q_k
=
\frac{1}{2n}\Bigl(\frac{p_k}{\tilde{E}}\Bigr)^{\frac{2+d}{2}},
\]
we obtain
\begin{align*}
R_2
&\leqslant
3\log n\cdot \tilde{E}^{-\frac{2+d}{2}}
\sum_{k=1}^{H-1} p_k^{\frac{4+d}{2}}\,\Delta_k
+
n\,\E_X\!\left[
|\Delta(X)|\,\mathbf{1}_{\{|\Delta(X)|<3\log n\cdot \Delta_H\}}
\right]
+ \E_X[|\Delta(X)|].
\end{align*}
By the definition of the instance complexity measure
\[
H(\nu)
=
\E_X\!\left[|\Delta(X)|^{\frac{2}{4+d}}\right]^{\frac{4+d}{2}},
\]
and using the same truncation argument as in the first-phase analysis to control the small-gap region, the right-hand side is bounded, up to logarithmic factors, by
\[
R_2
\;\leqslant\;
\softO\!\left(
\tilde{E}^{-\frac{2+d}{2}}\,H(\nu)
\;\vee\;
R_{\min}
\right).
\]

\medskip
\textbf{Conclusion.}
Combining the bounds for Phase~I and Phase~II, we conclude that, under the event $\mathcal{E}$,
the total regret satisfies
\[
R_{n}(\pi)
=
R_1+R_2
\;\leqslant\;
\softO\!\left(
\tilde{E}^{-\frac{2+d}{2}}\,H(\nu)
\;\vee\;
R_{\min}
\right),
\]
as claimed.

\paragraph{Estimation error.}
We now bound the estimation error of the final CATE estimator produced by the algorithm.

\medskip
\textbf{A sampling regularity event.}
Define an additional “good event” $\mathcal{E}_e$ under which each refined sub-cube receives a well-controlled number of samples from the optimal arm during Phase~II.
Specifically, let
\[
m := n-(r+r^2+\cdots+r^H),
\]
denote the length of Phase~II, so that $\tfrac{n}{2}\leqslant m\leqslant n$.
For each level-$k$ cube, recall that it is subdivided into $M_k^d$ sub-cubes, and the algorithm samples the optimal arm with probability $q_k$ on level-$k$ cubes.
We define
\begin{align*}
\mathcal{E}_e
:=
\Bigl\{
\text{for every level-$k$ cube and every one of its sub-cubes,}\\
\text{the sub-optimal arm is sampled }
N_{k,S}\in\Bigl[\tfrac{m a_k^d q_k}{2M_k^d},\,\tfrac{3m a_k^d q_k}{2M_k^d}\Bigr]
\Bigr\}.
\end{align*}

\medskip
\textbf{Probability of $\mathcal{E}_e$.}
Conditional on the covariates, the number of optimal-arm samples $N_{k,S}$ in a fixed sub-cube follows a binomial distribution with mean $\E[N_{k,S}]=m a_k^d q_k/M_k^d$.
Applying a Chernoff bound yields
\[
\mathbb{P}\!\left(
N_{k,S}\notin
\Bigl[\tfrac{1}{2}\E[N_{k,S}],\,\tfrac{3}{2}\E[N_{k,S}]\Bigr]
\right)
\leqslant
2\exp\!\left(-\tfrac{1}{12}\E[N_{k,S}]\right).
\]
Taking a union bound over all sub-cubes at all levels gives
\begin{align*}
\mathbb{P}(\mathcal{E}_e^c)
&\leqslant
\sum_{k=1}^H C_k M_k^d \cdot
2\exp\!\left(-\tfrac{1}{12} m a_k^d q_k/M_k^d\right) \\
&=
2\sum_{k=1}^H
p_k\,(n q_k)^{\frac{d}{2+d}}
\exp\!\left(-\tfrac{1}{12}(n q_k)^{\frac{2}{2+d}}\right),
\end{align*}
where we used $M_k^d\asymp (n q_k)^{\frac{d}{2+d}}$ and $m\asymp n$.
By the choice of $q_k$ and the fact that $H\asymp \log n$, the right-hand side is bounded by
\[
2(\log n)^d \exp\!\left(-\tfrac{1}{48}(\log n)^2\right),
\]
which is smaller than $1/n$ for all sufficiently large $n$.
Therefore,
\[
\mathbb{P}(\mathcal{E}_e)\;\geqslant\;1-\frac{1}{n}.
\]

Combining this with the earlier bound on $\mathcal{E}$ yields
\[
\mathbb{P}(\mathcal{E}\cap\mathcal{E}_e)\;\geqslant\;1-\frac{2}{n}.
\]

\medskip
\textbf{Bias--variance decomposition on a sub-cube.}
Fix a sub-cube $S$ obtained by refining a level-$k$ cube, and consider the estimator $\hat{\Delta}(S)$ formed by the difference of empirical means within $S$.
For a point $X\in S$, the mean squared error admits the standard decomposition
\begin{align*}
\E\!\left[(\hat{\Delta}(S)-\Delta(X))^2\right]
&=
\E\!\left[(\hat{\Delta}(S)-\Delta(S))^2\right]
+
\E\!\left[(\Delta(S)-\Delta(X))^2\right],
\end{align*}
corresponding to the variance and squared bias terms, respectively.

\medskip
\textbf{Control of bias and variance.}
On the event $\mathcal{E}\cap\mathcal{E}_e$, both terms can be controlled uniformly.
By Lipschitz continuity of $\Delta(\cdot)$ and the fact that $S$ has side length $a_k/M_k$, the squared bias satisfies
\begin{align*}
\text{bias}^2
&\leqslant
L^2\Bigl(\sqrt{d}\,\frac{a_k}{M_k}\Bigr)^2
\;\asymp\;
(n q_k)^{-\frac{2}{2+d}}.
\end{align*}
Moreover, under $\mathcal{E}_e$, the number of each arm samples in $S$ is at least $\tfrac{m a_k^d q_k}{2M_k^d}$.
Since the noise is $\sigma$-sub-Gaussian, the variance of the empirical mean difference is therefore bounded by
\begin{align*}
\text{variance}
&\leqslant
\sigma^2\Bigl(\tfrac{m a_k^d q_k}{2M_k^d}\Bigr)^{-1}
\;\asymp\;
(n q_k)^{-\frac{2}{2+d}}.
\end{align*}

\medskip
\textbf{Aggregation over all sub-cubes.}
Integrating over $X$ and summing over all levels, we obtain the overall estimation error bound
\begin{align*}
E
&\leqslant
\softO\!\left(
\sum_{k=1}^H p_k (n q_k)^{-\frac{2}{2+d}}
+ \mathbb{P}(\mathcal{E}^c\cup\mathcal{E}_e^c)
\right) \\
&\leqslant
\softO\!\left(
\tilde{E}
\;\vee\;
n^{-\frac{2}{2+d}}
\right).
\end{align*}
Since $\tilde{E}\geqslant n^{-\frac{2}{2+d}}$ by construction, this yields
\[
E
=
\softO\!\left(
\tilde{E}
\right),
\]
which completes the estimation error analysis.

\paragraph{Proof of $\hat{E}_{\max}\asymp\inf_{\nu'\in \mathcal{P}(\nu)} E_{\max}$.}
We show that the proposed estimator $\hat{E}_{\max}$ is order-wise equivalent to the smallest achievable
$E_{\max}$ among all instances sharing the same adaptive partition.

\medskip
\textbf{Unidentifiable small-gap region.}
Recall that gaps satisfying $|\Delta(X)|\leqslant n^{-\frac{1}{d+2}}$ cannot be reliably distinguished within
$n$ rounds.
Consequently, the quantity
\[
E_{\max}(\nu)
=
R_{\min}^{-\frac{2}{2+d}}(\nu)\,H(\nu)^{\frac{2}{2+d}}\ \vee\ E_{\min}
\]
may itself be unidentifiable from data, depending on the mass of this small-gap region.

We therefore consider the equivalence class
\[
\mathcal{P}(\nu)
:=
\{\nu' : \mathcal{P}_{\nu'}=\mathcal{P}_\nu\},
\]
consisting of all instances that induce the same adaptive partition as $\nu$
(see Definition~\ref{def: Adaptive Partition}).
Within this class, only the distribution of gaps inside the finest, statistically unresolvable region
may vary.

\medskip
\textbf{Decomposition of $E_{\max}$.}
By definition,
\begin{align*}
E_{\max}
&=
R_{\min}^{-\frac{2}{2+d}}(\nu)\,H(\nu)^{\frac{2}{2+d}}\ \vee\ E_{\min} \\
&=
\left(
\frac{
\Bigl(
\E_X\!\left[|\Delta(X)|^{\frac{2}{4+d}}\mathbf{1}\{|\Delta(X)|>n^{-\frac{1}{d+2}}\}\right]
+
\E_X\!\left[|\Delta(X)|^{\frac{2}{4+d}}\mathbf{1}\{|\Delta(X)|\leqslant n^{-\frac{1}{d+2}}\}\right]
\Bigr)^{\frac{4+d}{2}}
}{
\E_X\!\left[|\Delta(X)|^{-1-d}\mathbf{1}\{|\Delta(X)|>n^{-\frac{1}{d+2}}\}\right]
+
\E_X\!\left[n|\Delta(X)|\mathbf{1}\{|\Delta(X)|\leqslant n^{-\frac{1}{d+2}}\}\right]
}
\right)^{\frac{2}{2+d}}
\ \vee\ E_{\min}.
\end{align*}
Using the adaptive partition, the contribution from the identifiable region
$\{|\Delta(X)|>n^{-\frac{1}{d+2}}\}$ can be expressed in terms of partition statistics:
\[
\E_X\!\left[|\Delta(X)|^{\frac{2}{4+d}}\mathbf{1}\{|\Delta(X)|>n^{-\frac{1}{d+2}}\}\right]
\asymp
\sum_{k<H} p_k \Delta_k^{\frac{2}{4+d}},
\quad
\E_X\!\left[|\Delta(X)|^{-1-d}\mathbf{1}\{|\Delta(X)|>n^{-\frac{1}{d+2}}\}\right]
\asymp
\sum_{k<H} p_k \Delta_k^{-1-d}.
\]
Thus,
\begin{align*}
E_{\max}
\asymp
\left(
\frac{
\sum_{k<H} p_k\Delta_k^{\frac{2}{4+d}}
\ \vee\
\E_X\!\left[|\Delta(X)|^{\frac{2}{4+d}}\mathbf{1}\{|\Delta(X)|\leqslant n^{-\frac{1}{d+2}}\}\right]
}{
\sum_{k<H} p_k\Delta_k^{-1-d}
\ \vee\
\E_X\!\left[n|\Delta(X)|\mathbf{1}\{|\Delta(X)|\leqslant n^{-\frac{1}{d+2}}\}\right]
}
\right)^{\frac{2}{2+d}}
\ \vee\ E_{\min}.
\end{align*}

\medskip
\textbf{Worst-case choice within $\mathcal{P}(\nu)$.}
Since the function $x\mapsto x^{\frac{2}{4+d}}$ is concave, the smallest possible value of
$E_{\max}$ over $\nu'\in\mathcal{P}(\nu)$ is achieved by concentrating the unidentified mass
$p_H\asymp\mathbb{P}(|\Delta(X)|\leqslant n^{-\frac{1}{d+2}})$ at the boundary scale $0$ and $n^{-\frac{1}{d+2}}$. That is, $|\Delta(X)|$ in the part of $|\Delta(X)|\leqslant n^{-\frac{1}{d+2}}$ follows a two point distribution $(0, n^{-\frac{1}{d+2}})$ with the probability $(p,1-p)$ for a parameter $p\in [0,p_H]$.

Hence,
\begin{align*}
\inf_{\nu'\in\mathcal{P}(\nu)} E_{\max}
&\asymp
\inf_{p\in[0,p_H]}
\left(
\frac{
\sum_{k<H} p_k\Delta_k^{\frac{2}{4+d}}
\ \vee\
p^{\frac{4+d}{2}}
}{
\sum_{k<H} p_k\Delta_k^{-1-d}
\ \vee\
np
}
\right)^{\frac{2}{2+d}}
\ \vee\ E_{\min}.
\end{align*}
Define
\[
\Gamma_1 := n^{\frac{1}{2+d}}\sum_{k<H} p_k\Delta_k^{\frac{2}{4+d}},
\qquad
\Gamma_2 := n^{\frac{1}{2+d}}\sum_{k<H} p_k\Delta_k^{-1-d}.
\]
Then the preceding display can be written as
\[
\inf_{p\in[0,p_H]}
\left(
\frac{\Gamma_1\ \vee\ p^{\frac{4+d}{2}}}{\Gamma_2\ \vee\ np}
\right)^{\frac{2}{2+d}}
\ \vee\ E_{\min}.
\]

\medskip
\textbf{Reduction of the infimum.}
We first observe that
\begin{align*}
\Gamma_1^{\frac{2}{4+d}}
&=
n^{\frac{2}{(2+d)(4+d)}}\sum_{k<H} p_k\Delta_k^{\frac{2}{4+d}} \\
&\geqslant
n^{\frac{2}{(2+d)(4+d)}}\Bigl(\sum_{k<H} p_k\Bigr)
\left(n^{-\frac{1}{d+2}}\right)^{\frac{2}{4+d}}\\
&=
\sum_{k<H} p_k.
\end{align*}
Moreover,
\begin{align*}
\Gamma_2
&=
n^{\frac{1}{2+d}}\sum_{k<H} p_k\Delta_k^{-1-d}\\
&\leqslant \sum_{k<H} p_k\\
&\leqslant
n\,\Gamma_1^{\frac{2}{4+d}},
\end{align*}
which implies $\Gamma_1\geqslant \Gamma_2/n$.
As a result, the function
\[
p\mapsto
\left(
\frac{\Gamma_1\ \vee\ p^{\frac{4+d}{2}}}{\Gamma_2\ \vee\ np}
\right)^{\frac{2}{2+d}}
\]
is minimized at $p=\min\{p_H,\; \Gamma_1^{\frac{2}{4+d}}\}$, yielding
\[
\inf_{p\in[0,p_H]}
\left(
\frac{\Gamma_1\ \vee\ p^{\frac{4+d}{2}}}{\Gamma_2\ \vee\ np}
\right)^{\frac{2}{2+d}}
=
\left(
\frac{\Gamma_1}{\Gamma_2\ \vee\ np_H}
\right)^{\frac{2}{2+d}}.
\]

\medskip
\textbf{Conclusion.}
By definition, $\hat{E}_{\max}$ equals the right-hand side above up to logarithmic factors.
Therefore,
\[
\hat{E}_{\max}\;\asymp\;\inf_{\nu'\in\mathcal{P}(\nu)} E_{\max}.
\]
Furthermore, in the regimes of $p_H\leqslant \Gamma_2/n$ or $p_H\geqslant \Gamma_1$, the expression for
$E_{\max}(\nu')$ becomes independent of $\nu'\in\mathcal{P}(\nu)$.
In these cases, $\hat{E}_{\max}\asymp E_{\max}$ holds uniformly, completing the proof.

\section{Appendix: Proof of Theorem \ref{thm_upper_privacy}}
Throughout this appendix we assume the reward noise is $\sigma$-sub-Gaussian for all $(x,a)\in\X\times\{1,2\}$, and rewards are almost surely in $[0,1]$ (so that all $\ell_1$-sensitivities below are well-defined and bounded by constants). We prove (i) the privacy guarantee, and (ii) the same regret--estimation upper bounds as in Theorem~\ref{thm_upper_noprivacy}, up to $\softO(\cdot)$ factors, by following the same steps while tracking the additional Laplace perturbations.

\paragraph{Privacy guarantee (anticipating / joint DP).}
We use the anticipating (joint) differential privacy notion stated in Definition~\ref{def:JDP-interactive} (equivalently ADP/JDP): for any two neighboring datasets $D,D'$ differing at a single round $t$, and any measurable set $E\subseteq\{1,2\}^{n-1}$,
\[
\Pr(\,a_{-t}\in E \mid D\,)\ \leqslant\ e^\varepsilon \Pr(\,a_{-t}\in E \mid D'\,) ,
\]
where $a_{-t}=(a_1,\ldots,a_{t-1},a_{t+1},\ldots,a_n)$ and the probability is over the algorithm's internal randomness.

We prove that DP-ConSE is $\varepsilon$-ADP by isolating the only random releases that depend on any single individual and invoking post-processing.

\medskip
\textbf{Step 1: Phase I releases are $\varepsilon$-DP w.r.t.\ any single round in Phase I.}
Fix a batch index $i$ in Phase~I, and consider a fixed cube $S$ that is \textsc{NonTerminal} at the end of batch $i$.
Let $\hat\mu_{i,a}(S)$ be the empirical mean reward of arm $a$ using \emph{only} samples from batch $i$ whose covariates fall in $S$.
Define the (non-private) batch-$i$ empirical gap
\[
\hat\Delta_i^{\,0}(S)\ :=\ \hat\mu_{i,2}(S)-\hat\mu_{i,1}(S).
\]
DP-ConSE releases instead
\[
\hat\Delta_i(S)\ :=\ \hat\Delta_i^{\,0}(S)\ +\ Z_{i,S},
\qquad
Z_{i,S}\sim \text{Lap}\!\Big(\frac{2}{r^i\varepsilon}\Big),
\]
independently across $(i,S)$.

Now consider two neighboring datasets $D,D'$ differing at a single round $t$ that belongs to batch $i$.
That round contributes to \emph{exactly one} cube $S_t$ at level $i$ (the unique cube containing $X_t$), and to exactly one arm mean inside that cube (the arm actually pulled at time $t$).
Therefore, among all released statistics $\{\hat\Delta_i(S): S\in\mathcal P\}$ at the end of batch $i$, only \emph{one} coordinate (namely $S=S_t$) can change when we move from $D$ to $D'$; all others are computed from disjoint subsets of samples and are unchanged (parallel composition).

It remains to bound the $\ell_1$-sensitivity of $\hat\Delta_i^{\,0}(S_t)$ to changing one sample in batch $i$.
Within a fixed cube, changing a single reward in $[0,1]$ changes the corresponding empirical mean by at most $\frac{2}{r^i}$. Therefore, the Laplace mechanism with scale $2/(r^i\varepsilon)$ ensures that the release $\hat\Delta_i(S_t)$ is $\varepsilon$-DP with respect to the single differing round in batch $i$.
By parallel composition across cubes, the \emph{entire vector} $\{\hat\Delta_i(S)\}_{S}$ released at the end of batch $i$ is $\varepsilon$-DP with respect to any single change within batch $i$.

\medskip
\textbf{Step 2: Phase II releases are $\varepsilon$-DP w.r.t.\ any single round in Phase II.}
In Phase~II, for each refined sub-cube $B$ and arm $a$, the algorithm forms the (non-private) sufficient statistics:
\[
s_a(B)=\sum_{t\in \text{Phase II}}\mathbf 1\{X_t\in B,\ a_t=a\}\,Y_t,
\qquad
c_a(B)=\sum_{t\in \text{Phase II}}\mathbf 1\{X_t\in B,\ a_t=a\},
\]
and releases the privatized mean
\[
\tilde\mu_a(B)\ :=\ \frac{s_a(B)+U_{a,B}}{(c_a(B)+V_{a,B})\vee 1},
\qquad
U_{a,B},V_{a,B}\ \sim\ \text{Lap}\!\Big(\frac{2}{\varepsilon}\Big),
\]
independently across $(a,B)$.

Fix two neighboring datasets $D,D'$ differing at a single Phase~II round $t$.
That round belongs to exactly one sub-cube $B_t$ and exactly one arm $a_t$.
Hence \emph{only} $(s_{a_t}(B_t),c_{a_t}(B_t))$ can change; all other pairs $(s_a(B),c_a(B))$ are unchanged (parallel composition across $(a,B)$).
Moreover, changing one record changes $c_{a_t}(B_t)$ by at most $1$ and changes $s_{a_t}(B_t)$ by at most $1$ (since $Y_t\in[0,1]$), so both have $\ell_1$-sensitivity at most $1$.
The Laplace mechanism with scale $2/\varepsilon$ therefore makes each of the releases $s_{a_t}(B_t)+U_{a_t,B_t}$ and $c_{a_t}(B_t)+V_{a_t,B_t}$ individually $(\varepsilon/2)$-DP, and by (sequential) composition the pair is $\varepsilon$-DP.
Finally, $\tilde\mu_{a_t}(B_t)$ is a deterministic post-processing of these two released values, so it is also $\varepsilon$-DP.

\medskip
\textbf{Step 3: From private releases to private actions (ADP).}
Fix a neighboring change at time $t$.
All actions prior to time $t$ are causally independent of $(X_t,Y_t)$ and hence have identical distributions under $D$ and $D'$.
All actions after time $t$ are computed as a (possibly adaptive) function of:
(i) the public covariates,
(ii) internal randomness,
and (iii) the \emph{released} noisy statistics from Phase~I/II.
By Steps 1--2, the collection of released statistics that can be influenced by the single change at time $t$ is $\varepsilon$-DP.
Therefore, by the post-processing property, the induced distribution of future actions $(a_{t+1},\ldots,a_n)$ is $\varepsilon$-DP with respect to that change.
Equivalently, the full action sequence excluding user $t$, namely $a_{-t}$, satisfies $\varepsilon$-ADP.

This completes the privacy proof.

\paragraph{A good event for Laplace noises.}
We next control the added Laplace noises uniformly.

\medskip
\textbf{Phase I Laplace noises.}
For each pair $(i,S)$ where the algorithm releases $\hat\Delta_i(S)$, the added noise is $Z_{i,S}\sim\text{Lap}(2/(r^i\varepsilon))$, so
\[
\Pr\big(|Z_{i,S}|>u\big)\ =\ \exp\!\Big(-\frac{u r^i\varepsilon}{2}\Big).
\]
Set $u_{i}:=\frac{4\log n}{r^i\varepsilon}$ so that $\Pr(|Z_{i,S}|>u_i)=n^{-2}$.
Define
\[
\mathcal E_{\varepsilon,1}
:=
\Bigl\{
\forall\,1\leqslant i\leqslant H,\ \forall\,S,\ |Z_{i,S}|\leqslant \tfrac{4\log n}{r^i\varepsilon}
\Bigr\}.
\]
Since at step $i$ there are at most $2^{di}$ cubes, the total number of released $(i,S)$ pairs is at most
$\sum_{i=1}^H 2^{di}\leqslant H2^{dH}\leqslant \frac{2}{2+d}\log_2 n\cdot n^{\frac{d}{2+d}}$.
A union bound gives, for all sufficiently large $n$,
\[
\Pr(\mathcal E_{\varepsilon,1}^c)
\leqslant n^{-2}\cdot \frac{2}{2+d}\log_2 n\cdot n^{\frac{d}{2+d}}
\leqslant \frac{1}{2n},
\qquad
\Pr(\mathcal E_{\varepsilon,1})\geqslant 1-\frac{1}{2n}.
\]

\medskip
\textbf{Phase II Laplace noises.}
For each released pair $(U_{a,B},V_{a,B})$ with $U_{a,B},V_{a,B}\sim\text{Lap}(2/\varepsilon)$, define
\[
\mathcal E_{\varepsilon,2}
:=
\Bigl\{
\forall\,a\in\{1,2\},\ \forall\,B,\ |U_{a,B}|\leqslant \tfrac{4\log n}{\varepsilon},\ |V_{a,B}|\leqslant \tfrac{4\log n}{\varepsilon}
\Bigr\}.
\]
The number of released $(a,B)$ pairs is at most $2\sum_{k=0}^H C_k M_k^d$, which is polynomial in $n$ and at most $n$ up to logarithmic factors under the algorithm's choice of $M_k$.
Since $\Pr(|\text{Lap}(2/\varepsilon)|>4\log n/\varepsilon)=n^{-2}$, another union bound yields
\[
\Pr(\mathcal E_{\varepsilon,2})\geqslant 1-\frac{1}{2n}
\]
for all sufficiently large $n$.

\paragraph{A modified high-probability event for empirical gaps.}
Define the (non-private) empirical gap $\hat\Delta_i^{\,0}(S)$ and the released gap $\hat\Delta_i(S)=\hat\Delta_i^{\,0}(S)+Z_{i,S}$.
Let $\Delta(S)$ denote the true average gap over cube $S$.
We define a “good event” ensuring uniform accuracy of the \emph{released} gaps:
\[
\mathcal E
:=
\Bigl\{
\forall\,1\leqslant i\leqslant H,\ \forall\,S,\ 
\bigl|\Delta(S)-\hat\Delta_i(S)\bigr|
\leqslant
\underbrace{2\sigma\sqrt{\log n}\,\Delta_i}_{\text{sub-Gaussian term}}
+
\underbrace{\tfrac{4\log n}{r^i\varepsilon}}_{\text{Laplace term}}
\Bigr\}.
\]
Fix $(i,S)$.
Conditioning on the covariates and actions inside batch $i$, $\hat\Delta_i^{\,0}(S)$ is an average of independent $\sigma$-sub-Gaussian noise terms at scale $\Delta_i$, hence
\[
\Pr\!\left(\bigl|\Delta(S)-\hat\Delta_i^{\,0}(S)\bigr|>2\sigma\sqrt{\log n}\,\Delta_i\right)\leqslant \frac{2}{n^2}.
\]
Also $\Pr(|Z_{i,S}|>4\log n/(r^i\varepsilon))=n^{-2}$.
By a union bound over these two deviations,
\[
\Pr\!\left(\bigl|\Delta(S)-\hat\Delta_i(S)\bigr|>
2\sigma\sqrt{\log n}\,\Delta_i+\tfrac{4\log n}{r^i\varepsilon}\right)
\leqslant \frac{3}{n^2}.
\]
Taking a union bound over all $(i,S)$ pairs (at most $H2^{dH}$ of them) yields, for all sufficiently large $n$,
\[
\Pr(\mathcal E)\ \geqslant\ 1-\frac{1}{n}.
\]
In the remainder of the proof we condition on the event
\[
\mathcal G\ :=\ \mathcal E\cap \mathcal E_{\varepsilon,1}\cap \mathcal E_{\varepsilon,2},
\qquad
\Pr(\mathcal G)\geqslant 1-\frac{2}{n}.
\]

\paragraph{Regret analysis under $\mathcal G$: Phase I.}
The regret argument follows the proof of Theorem~\ref{thm_upper_noprivacy}; we only track the modified threshold and verify that Laplace noise does not change the order.

\medskip
\textbf{Correctness of terminal labels.}
A cube $S$ at step $k$ is declared terminal if $|\hat\Delta_k(S)|>(\log n)^2\Delta_k$ (with sign determining the labeled optimal arm).
Under $\mathcal G$, the released estimate $\hat\Delta_k(S)$ differs from $\Delta(S)$ by at most
$2\sigma\sqrt{\log n}\Delta_k + 4\log n/(r^k\varepsilon)$.
For all sufficiently large $n$, the Laplace term is negligible compared to $(\log n)^2\Delta_k$ because
$r^k=2^{(d+2)k}=(2^k)^{d+2}=\Delta_k^{-(d+2)}$ implies
\[
\frac{4\log n}{r^k\varepsilon}\ =\ \frac{4\log n}{\varepsilon}\,\Delta_k^{d+2}
\ \ll\ (\log n)^2\Delta_k
\quad \text{uniformly for } k\leqslant H.
\]
Hence, whenever $S$ is declared terminal, the labeled arm is truly optimal on $S$.
Thus, in Phase~I regret is incurred only in cubes that are not yet terminal, where both arms are explored.

\medskip
\textbf{Gap control on non-terminal cubes.}
Fix step $k$ and a cube $S$ that is not terminal.
Then $|\hat\Delta_k(S)|\leqslant (\log n)^2\Delta_k$.
On $\mathcal G$,
\begin{align*}
|\Delta(S)|
&\leqslant |\hat\Delta_k(S)|+\bigl|\Delta(S)-\hat\Delta_k(S)\bigr|\\
&\leqslant (\log n)^2\Delta_k + 2\sigma\sqrt{\log n}\,\Delta_k + \frac{4\log n}{r^k\varepsilon}\\
&\leqslant \frac{3}{2}(\log n)^2\Delta_k,
\end{align*}
for all sufficiently large $n$, absorbing lower-order terms into $(\log n)^2$.

If $S$ has side length $a_k=2^{-k}=\Delta_k$, Lipschitz continuity of $\Delta(\cdot)$ implies that for any $x\in S$,
\begin{align*}
|\Delta(x)|
&\leqslant |\Delta(S)| + L\|x-x_S\|\\
&\leqslant |\Delta(S)| + L\sqrt d\,a_k\\
&\leqslant 2(\log n)^2\Delta_k,
\end{align*}
again absorbing constants into $(\log n)^2$.

\medskip
\textbf{Termination depth and Phase I regret.}
Define
\[
k_x:=\min\bigl\{k:\ |\Delta(x)|>2(\log n)^2\Delta_k\bigr\}.
\]
Since $\Delta_k=2^{-k}$, we obtain
\[
k_x
=
\Bigl\lfloor
\log_2\bigl((\log n)^2\bigr) + 1 - \log_2|\Delta(x)|
\Bigr\rfloor+1.
\]
With $r=2^{d+2}$, this yields
\[
r^{k_x-1}
\leqslant
\bigl(4\log n\bigr)^{2+d}\,|\Delta(x)|^{-(2+d)}.
\]
As in Theorem~\ref{thm_upper_noprivacy}, the number of exploratory pulls at covariate $x$ before termination is at most
$r+\cdots+r^{k_x}\leqslant r^2\min\{r^{k_x-1},n\}$.
Therefore, accounting for $\Pr(\mathcal G^c)\leqslant 2/n$,
\begin{align*}
R_1
&\leqslant
r^2\int_{\X} |\Delta(x)|\min\{r^{k_x-1},n\}\,d\mathcal P(x)
+\Pr(\mathcal G^c)\cdot \frac{n}{2}\E[|\Delta(X)|]\\
&\leqslant
\softO\!\left(
\E_X\!\left[\min\{|\Delta(X)|^{-(1+d)},\ n|\Delta(X)|\}\right]
\right)
=:R_{\min}.
\end{align*}

\paragraph{Regret analysis under $\mathcal G$: Phase II.}
The Phase~II regret argument is identical to Theorem~\ref{thm_upper_noprivacy} once we have upper/lower bounds on $|\Delta(x)|$ within a terminal cube, which remain valid under $\mathcal G$.

\medskip
\textbf{Gap magnitude on terminal cubes.}
If a cube $S$ terminates at step $k<H$, then $|\hat\Delta_k(S)|>(\log n)^2\Delta_k$.
For any $x\in S$, using $\mathcal G$ and Lipschitz continuity,
\begin{align*}
|\Delta(x)|
&\geqslant |\Delta(S)|-L\sqrt d\,a_k\\
&\geqslant |\hat\Delta_k(S)|-\bigl|\Delta(S)-\hat\Delta_k(S)\bigr|-L\sqrt d\,\Delta_k\\
&\geqslant (\log n)^2\Delta_k - 2\sigma\sqrt{\log n}\,\Delta_k - \frac{4\log n}{r^k\varepsilon}-L\sqrt d\,\Delta_k\\
&\geqslant \frac{1}{2}(\log n)^2\Delta_k,
\end{align*}
for all sufficiently large $n$.

Let $S'$ be the parent cube of $S$ at level $k-1$.
Since $S'$ did not terminate at step $k-1$, we have $|\hat\Delta_{k-1}(S')|\leqslant (\log n)^2\Delta_{k-1}=2(\log n)^2\Delta_k$.
Applying $\mathcal G$ and Lipschitz continuity yields, for any $x\in S$,
\begin{align*}
|\Delta(x)|
&\leqslant |\Delta(S')|+L\sqrt d\,a_{k-1}\\
&\leqslant |\hat\Delta_{k-1}(S')| + \bigl|\Delta(S')-\hat\Delta_{k-1}(S')\bigr| + L\sqrt d\,\Delta_{k-1}\\
&\leqslant 2(\log n)^2\Delta_k + 2\sigma\sqrt{\log n}\,\Delta_{k-1}+\frac{4\log n}{r^{k-1}\varepsilon}+L\sqrt d\,\Delta_{k-1}\\
&\leqslant 3(\log n)^2\Delta_k,
\end{align*}
absorbing lower-order terms into $(\log n)^2$.
The same upper bound holds for cubes that remain non-terminal until the final step $H$.

\medskip
\textbf{Phase II regret bound.}
During Phase~II, on a level-$k$ terminal cube ($k<H$), the algorithm exploits the labeled optimal arm with probability $1-q_k$ and randomizes with probability $q_k$.
Each randomized pull incurs instantaneous regret at most $|\Delta(x)|\leqslant 3(\log n)^2\Delta_k$ on $\mathcal G$.
Therefore,
\begin{align*}
R_2
&\leqslant
n\Bigg(
\sum_{k=1}^{H-1} p_k q_k\cdot 3(\log n)^2\Delta_k
+
\E_X\!\left[|\Delta(X)|\mathbf 1\{|\Delta(X)|<3(\log n)^2\Delta_H\}\right]
\Bigg)
+\Pr(\mathcal G^c)\cdot n\E[|\Delta(X)|].
\end{align*}
Using the same choice of $q_k$ as in the non-private algorithm and the same truncation argument controlling the small-gap region, we conclude
\[
R_2
\leqslant
\softO\!\left(\tilde E^{-\frac{2+d}{2}}H(\nu)\ \vee\ R_{\min}\right).
\]
Combining with $R_1\leqslant R_{\min}$ gives the claimed total regret bound.

\paragraph{Estimation error under $\mathcal G$.}
We now bound the final mean-squared CATE error. The proof follows Theorem~\ref{thm_upper_noprivacy}, with one additional term controlling the perturbation introduced by releasing privatized means.

\medskip
\textbf{Sampling regularity event.}
Let $\mathcal E_e$ be the same Chernoff-type event as in Theorem~\ref{thm_upper_noprivacy} ensuring that each refined sub-cube receives the correct order of samples from the (labeled) optimal arm in Phase~II:
\[
\mathcal E_e
:=
\Bigl\{
\forall k,\forall B\subset S\text{ (a refined sub-cube at level }k),
\ N_{k,B}\in\bigl[\tfrac{m a_k^d q_k}{2M_k^d},\tfrac{3m a_k^d q_k}{2M_k^d}\bigr]
\Bigr\},
\]
where $m$ is the Phase~II length, and $N_{k,B}$ is the number of Phase~II samples in $B$ from the arm used for estimation there.
As in Theorem~\ref{thm_upper_noprivacy},
\[
\Pr(\mathcal E_e)\geqslant 1-\frac{1}{n}
\quad\text{for all sufficiently large }n.
\]
Hence $\Pr(\mathcal G\cap\mathcal E_e)\geqslant 1-\frac{3}{n}$.

\medskip
\textbf{Bias--variance--privacy decomposition on a sub-cube.}
Fix a refined sub-cube $B$ at level $k$, and a point $X\in B$.
Let $\hat\mu_a^{\,0}(B)=s_a(B)/c_a(B)$ be the non-private empirical mean and $\tilde\mu_a(B)$ the released private mean.
Define the private gap estimator on $B$ by $\tilde\Delta(B):=\tilde\mu_2(B)-\tilde\mu_1(B)$ (and similarly $\Delta(B)$ for the true average gap on $B$).
Then
\begin{align*}
\E\!\left[(\tilde\Delta(B)-\Delta(X))^2\right]
&\leqslant
3\,\E\!\left[(\tilde\Delta(B)-\hat\Delta^{\,0}(B))^2\right]
+3\,\E\!\left[(\hat\Delta^{\,0}(B)-\Delta(B))^2\right]
+3\,\E\!\left[(\Delta(B)-\Delta(X))^2\right],
\end{align*}
corresponding respectively to (i) the DP perturbation error, (ii) statistical variance, and (iii) squared bias due to Lipschitz variation within $B$.

\medskip
\textbf{Squared bias and statistical variance (as before).}
On $\mathcal G\cap\mathcal E_e$, the sub-cube side length is $a_k/M_k$, so Lipschitz continuity yields
\[
\E\!\left[(\Delta(B)-\Delta(X))^2\right]\ \leqslant\ L^2\Bigl(\sqrt d\,\frac{a_k}{M_k}\Bigr)^2\ \asymp\ (nq_k)^{-\frac{2}{2+d}}.
\]
Also $c_a(B)\gtrsim m a_k^d q_k/M_k^d \asymp (nq_k)^{\frac{2}{2+d}}$, hence the variance of $\hat\mu_a^{\,0}(B)$ is $\lesssim \sigma^2/c_a(B)$ and
\[
\E\!\left[(\hat\Delta^{\,0}(B)-\Delta(B))^2\right]\ \lesssim\ (nq_k)^{-\frac{2}{2+d}}.
\]

\medskip
\textbf{DP perturbation error for the released ratio.}
We bound the deviation between $\tilde\mu_a(B)$ and $\hat\mu_a^{\,0}(B)$ on $\mathcal E_{\varepsilon,2}$.
Write $s:=s_a(B)$ and $c:=c_a(B)$.
Since $Y_t\in[0,1]$, we have $0\leqslant s\leqslant c$ and hence $|s|\leqslant c$.
On $\mathcal E_{\varepsilon,2}$, $|U_{a,B}|\leqslant 4\log n/\varepsilon$ and $|V_{a,B}|\leqslant 4\log n/\varepsilon$.
Assume additionally that $c\geqslant 8\log n/\varepsilon$ (this will hold under $\mathcal E_e$ for all sufficiently large $n$ because $c\asymp (nq_k)^{2/(2+d)}\gtrsim (\log n)^2$).
Then
\[
c+V_{a,B}\ \geqslant\ c-\frac{4\log n}{\varepsilon}\ \geqslant\ \frac{c}{2}.
\]
Therefore,
\begin{align*}
\left|\tilde\mu_a(B)-\hat\mu_a^{\,0}(B)\right|
&=
\left|\frac{s+U_{a,B}}{(c+V_{a,B})\vee 1}-\frac{s}{c}\right|
\ \leqslant\
\left|\frac{s+U_{a,B}}{c+V_{a,B}}-\frac{s}{c}\right|\\
&=
\left|\frac{(s+U_{a,B})c-s(c+V_{a,B})}{c(c+V_{a,B})}\right|
=
\left|\frac{U_{a,B}c-sV_{a,B}}{c(c+V_{a,B})}\right|\\
&\leqslant
\frac{|U_{a,B}|c+|s||V_{a,B}|}{c(c+V_{a,B})}
\ \leqslant\
\frac{|U_{a,B}|+|V_{a,B}|}{c+V_{a,B}}
\ \leqslant\
\frac{2(|U_{a,B}|+|V_{a,B}|)}{c}\\
&\leqslant
\frac{16\log n}{\varepsilon\,c}.
\end{align*}
Since $c\asymp (nq_k)^{\frac{2}{2+d}}$ on $\mathcal E_e$, this yields
\[
\E\!\left[(\tilde\mu_a(B)-\hat\mu_a^{\,0}(B))^2\right]
\ \leqslant\
\softO\!\left(\frac{1}{\varepsilon^2}\,(nq_k)^{-\frac{4}{2+d}}\right)
\ =\
\softO\!\left((nq_k)^{-\frac{2}{2+d}}\right),
\]
where the last step uses that $\varepsilon>0$ is fixed and $(nq_k)^{-\frac{4}{2+d}}$ is of strictly smaller order than $(nq_k)^{-\frac{2}{2+d}}$ (hence absorbed into $\softO(\cdot)$).

Combining the three components, we conclude that on $\mathcal G\cap\mathcal E_e$,
\[
\E\!\left[(\tilde\Delta(B)-\Delta(X))^2\right]\ \leqslant\ \softO\!\left((nq_k)^{-\frac{2}{2+d}}\right),
\]
uniformly over sub-cubes at level $k$.

\medskip
\textbf{Aggregation over levels.}
Integrating over $X$ and summing over levels as in Theorem~\ref{thm_upper_noprivacy}, we obtain
\begin{align*}
E
&\leqslant
\softO\!\left(
\sum_{k=1}^H p_k (nq_k)^{-\frac{2}{2+d}}
+\Pr\big((\mathcal G\cap\mathcal E_e)^c\big)
\right)
\ \leqslant\
\softO\!\left(\tilde E\right),
\end{align*}
since $\Pr((\mathcal G\cap\mathcal E_e)^c)\leqslant 3/n$ and $\tilde E\geqslant E_{\min}\gtrsim n^{-2/(2+d)}$.

\paragraph{Conclusion.}
We have shown:
(i) DP-ConSE satisfies $\varepsilon$-anticipating (joint) differential privacy;
(ii) under the high-probability event $\mathcal G\cap\mathcal E_e$, the regret and estimation analyses follow the non-private proof with only lower-order Laplace contributions, giving
\[
R_{n,\nu}\ =\ \softO\!\left(\tilde E^{-\frac{2+d}{2}}H(\nu)\ \vee\ R_{\min}\right),
\qquad
E_{n,\nu}\ =\ \softO(\tilde E).
\]
This completes the proof of Theorem~\ref{thm_upper_privacy}.

\section{Appendix: Proof of Theorem \ref{thm:simple_regret}} \label{proof: simple regret}
On the high-probability event $\mathcal{E}$, all cubes that are declared \textsc{Terminal} are labeled with the correct optimal arm. Consequently, the simple regret can arise only when the covariate $X$ falls into a cube that is not terminated before the final depth $H$. By the termination rule of the algorithm and the accuracy guarantees on $\mathcal{E}$, the gap on any such cube is uniformly bounded by
\[
2(\log n)^2\Delta_H
=
2(\log n)^2 n^{-\frac{1}{2+d}}.
\]
Therefore, decomposing according to whether $\mathcal{E}$ occurs, the simple regret is bounded by
\begin{align*}
\softO\left(
\frac{1}{n}
+
\E_X\left[
|\Delta(X)|
\mathbf{1}\left\{|\Delta(X)|\leqslant 2(\log n)^2 n^{-\frac{1}{2+d}}\right\}
\right]
\right),
\end{align*}
where the $\frac{1}{n}$ term accounts for the negligible contribution from the complement event $\mathcal{E}^c$.

This bound matches, up to logarithmic factors, the minimax rate established in \cite{simple-regret-instance-lowerbound} for the margin condition class $V(\alpha, 1)$ with $\alpha\leqslant d$, namely $n^{-\frac{1+\alpha}{2+d}}$. Indeed, for $\nu\in V(\alpha)$, we have
\begin{align*}
&\sup_{\nu\in V(\alpha)}
\left\{
\frac{1}{n}
+
\E_X\left[
|\Delta_\nu(X)|
\mathbf{1}\left\{|\Delta_\nu(X)|\leqslant 2(\log n)^2 n^{-\frac{1}{2+d}}\right\}
\right]
\right\}\\
&=
\frac{1}{n}
+
2(\log n)^2 n^{-\frac{1}{2+d}}
\,D\!\left(2(\log n)^2 n^{-\frac{1}{2+d}}\right)^{\alpha} \\
&=
\softO\!\left(n^{-\frac{1+\alpha}{2+d}}\right),
\end{align*}
which establishes the claimed rate.

\end{document}